\documentclass[twocolumn]{aastex701}
\usepackage{graphics,graphicx}
\hypersetup{urlcolor=blue}
\usepackage{natbib}
\usepackage[outdir=./]{epstopdf}
\usepackage{textcomp,gensymb}
\usepackage{xfrac}
\usepackage{xcolor}
\usepackage{multirow}
\usepackage{longtable}
\usepackage{adjustbox}

\newcommand{\mps}{m\,s$^{-1}$}

\newcommand{\vsini}{$v\sin{i_*}$}

\newcommand{\kepler}{{\it Kepler}}
\newcommand{\logg}{$log~g$ }

\newcommand{\um}{$\mu$m}
\newcommand{\fbol}{$F_{\mathrm{bol}}$}

\newcommand{\teff}{\ensuremath{T_{\text{eff}}}}
\newcommand\kms{km~s$^{-1}$}

\newcommand{\gaia}{{\textit Gaia}}

\usepackage[outdir=./]{epstopdf}
\usepackage{ulem}
\usepackage{mathrsfs}
\usepackage{fontawesome}
\usepackage{comment}
\usepackage[version=4]{mhchem}

\newcommand{\tess}{\textit{TESS}}
\newcommand{\ktwo}{{\textit K2}}

\newcommand{\maroonx}{\textit{MAROON-X}}
\newcommand{\starname}{TIC 150070085}
\newcommand{\planetname}{TIC 150070085 b}
\newcommand{\assoc}{Alessi 84}

\newcommand{\eso}{European Southern Observatory, Karl-Schwarzschildstraße 2, D-85748 Garching bei München, Germany}

\pdfoutput=1

\shorttitle{SOYSAUCE II} 
\shortauthors{Barber et al. }

\begin{document}

\title{Stellar Obliquities of Young Systems, Atmospheres Undergoing Contraction and Escape (SOYSAUCE) II: a 135\,Myr planet on an aligned orbit with transit timing variations}

\correspondingauthor{Madyson G. Barber}
\email{madysonb@live.unc.edu}

\author[0000-0002-8399-472X]{Madyson G. Barber}
\email{madysonb@live.unc.edu}
\altaffiliation{NSF Graduate Research Fellow}
\affiliation{Department of Physics and Astronomy, The University of North Carolina at Chapel Hill, Chapel Hill, NC 27599, USA} 

\author[0000-0003-3654-1602]{Andrew W. Mann}
\email{awmann@unc.edu}
\affiliation{Department of Physics and Astronomy, The University of North Carolina at Chapel Hill, Chapel Hill, NC 27599, USA}

\author[0000-0001-9158-9276]{Sydney Vach}
\affiliation{\eso}
\email{sydneyvach.astro@gmail.com}

\author[0009-0008-0940-1317]{Leah J. Boff}
\email{lboff@unc.edu}
\altaffiliation{UNC Chancellor's Science Scholar}
\affiliation{Department of Physics and Astronomy, The University of North Carolina at Chapel Hill, Chapel Hill, NC 27599, USA} 
\email{lboff@unc.edu}

\author[0000-0001-6037-2971]{Andrew W. Boyle}
\altaffiliation{NSF Graduate Research Fellow}
\affiliation{Department of Physics and Astronomy, The University of North Carolina at Chapel Hill, Chapel Hill, NC 27599, USA} 
\email{awboyle@unc.edu}  

\author[0000-0001-7246-5438]{Andrew Vanderburg}
\affiliation{Center for Astrophysics \textbar \ Harvard \& Smithsonian, 60 Garden Street, Cambridge, MA 02138, USA}
\email{andrew.m.vanderburg@gmail.com}

\author[0000-0001-9811-568X]{Adam L. Kraus}
\affiliation{Department of Astronomy, The University of Texas at Austin, Austin, TX 78712, USA}
\email{alk@astro.as.utexas.edu}

\author[0000-0003-2053-0749]{Benjamin M.\ Tofflemire}
\affiliation{SETI Institute, Mountain View, CA 94043, USA/NASA Ames Research Center, Moffett Field, CA 90345 USA}
\affiliation{Department of Astronomy, The University of Texas at Austin, Austin, TX 78712, USA}
\email{tofflemire@utexas.edu}

\author[0000-0002-5099-8185]{Marshall C. Johnson}
\affiliation{Department of Astronomy, The Ohio State University, Columbus, OH 43210, USA}
\email{johnson.7240@osu.edu}

\author[0000-0001-6637-5401]{Allyson Bieryla} 
\affiliation{Center for Astrophysics \textbar \ Harvard \& Smithsonian, 60 Garden Street, Cambridge, MA 02138, USA}
\email{abieryla@cfa.harvard.edu}

\author[0000-0001-9911-7388]{David~W.~Latham}
\affiliation{Center for Astrophysics \textbar \ Harvard \& Smithsonian, 60 Garden Street, Cambridge, MA 02138, USA}
\email{dlatham@cfa.harvard.edu}

\author[0000-0001-6588-9574]{Karen A.\ Collins}
\affiliation{Center for Astrophysics \textbar \ Harvard \& Smithsonian, 60 Garden Street, Cambridge, MA 02138, USA}
\email{karen.collins@cfa.harvard.edu}

\author[0000-0002-2532-2853]{Steve B. Howell}
\affiliation{NASA Ames Research Center, Moffett Field, CA 94035 USA}
\email{steve.b.howell@nasa.gov}

\author[0000-0001-8227-1020]{Richard P. Schwarz}
\affiliation{Center for Astrophysics \textbar \ Harvard \& Smithsonian, 60 Garden Street, Cambridge, MA 02138, USA}
\email{rpschwarz@comcast.net}

\author{Gregorg Srdoc}
\affil{Kotizarovci Observatory, Sarsoni 90, 51216 Viskovo, Croatia}
\email{gregorsrdoc@gmail.com}

\author[0000-0003-2127-8952]{Francis P. Wilkin} 
\affiliation{Department of Physics and Astronomy, Union College, 807 Union St., Schenectady, NY 12308, USA}
\email{wilkinf@union.edu}

\author[0000-0001-9087-1245]{Felipe Murgas}
\affiliation{Instituto de Astrof\'\i sica de Canarias (IAC), E-38205 La Laguna, Tenerife, Spain}
\affiliation{Departamento de Astrof\'\i sica, Universidad de La Laguna (ULL), E-38206, La Laguna, Tenerife, Spain}
\email{fmurgas@iac.es}

\author[0000-0003-0987-1593]{Enric Palle}
\affiliation{Instituto de Astrof\'\i sica de Canarias (IAC), E-38205 La Laguna, Tenerife, Spain}
\affiliation{Departamento de Astrof\'\i sica, Universidad de La Laguna (ULL), E-38206, La Laguna, Tenerife, Spain}
\email{epalle@iac.es}

\author[0000-0003-2163-1437]{Chris Stockdale}
\affiliation{Hazelwood Observatory, Churchill, Victoria, Australia}
\email{thestockdalefamily@bigpond.com}

\begin{abstract}

Young planets ($<$1\,Gyr) provide opportunities to directly probe planet formation and evolution processes in action. However, due to heightened stellar activity, there is a lack of known transiting planets in adolescence ($\sim$100-500\,Myr). Here we present the validation of \planetname, a 3.6 $R_\oplus$ planet on a 10.47 day orbit, and report the candidate \starname\,c, a 3.0 $R_\oplus$ planet on a 15.90 day orbit. While we are unable to validate the second signal, the proximity to mean motion resonance (3:2) and transit timing variations observed in the transits of \planetname\ strongly suggest the signal is planetary. We confirm the host star as a member of Alessi 84 and combine the group's CMD, rotation, and variability properties to update the age to $135\pm10$ Myr. We additionally use \maroonx\ to observe the Rossiter-McLaughlin signal of \planetname\ and measure the sky projected obliquity angle ($\lambda$). We find \planetname\ is consistent with a near-aligned orbit with its host star ($|\lambda| = 18\pm12\degree$), in line with similarly aged transiting planets with measured $\lambda$ values. Continued discovery and characterization of planets in this age regime are vital to link planetary infancy ($<$50\,Myr) and maturity ($>$1\,Gyr).

\end{abstract}


\section{Introduction} \label{sec:intro}


The study of young planetary systems ($<1$\,Gyr) offers a unique window into the processes that shape the architecture and physical characteristics of mature exoplanets. By observing planets shortly after their formation, we can directly test models of planet formation and evolution \citep{Izidoro2017, Rogers2025}, helping to determine which early conditions and physical mechanisms give rise to the diverse population of planets observed by \kepler{} and similar demographic surveys \citep{2013PNAS..11019273P, Clanton2016, Berger2020}. 

Planets found within stellar associations are particularly valuable for these studies because such groups provide the most accurate and precise stellar ages \citep{Soderblom2010}. Unlike field stars, members of associations allow for age determination through a suite of mutually reinforcing methods \citep{Barber2026_6448}, including isochronal modeling of pre-main-sequence stars \citep[e.g.,][]{Bell2015}, gyrochronology \citep{Curtis2019_stalling, Bouma2023_gyrointerp}, activity levels \citep{Barber2023}, lithium depletion \citep{Jeffries2023_eagles, Wood2023}, and kinematic traceback \citep{Boyle2025b, Couture2026}. 

Data from \ktwo\ and \tess\ have fueled an explosion in the discovery of young transiting planets. This includes newly-formed planets like IRAS~04125+2902 \citep[3\,Myr;][]{Barber2024_iras}, systems observed by {\it JWST} \citep[e.g.,][]{THYMEII,Thao2024_featherweight, David2019_v1298tau,Barat2025}, lava worlds \citep{Hedges2021}, and chains of well-packed multi-transiting systems \citep{Plavchan2020,David2019_v1298tau,Dattilo2025}. Such young planets have been critical for a range of science results, including the finding that young planets are larger than their older counterparts \citep{Fernandes2022, Vach2024_occ} and are preferentially found in or near resonant orbits \citep{Dai2024, LopezMurillo2026}. The abundance and properties of young planets also favors Type I disk-driven migration \citep{Lee2013, Izidoro2017, Ogihara2018}. 

As noted in the first paper of this series, SOYSAUCE I \citep{soysauce1}, there is tentative evidence that young planets tend to be more well-aligned with their host star's rotational axis than older populations \citep[e.g.,][]{Johnson2022, Dai2023, Teng2024}. The youngest known misaligned system is $\simeq$210 Myr (Kepler-63\,b), which shows an impressive true (3D) obliquity $\psi=114^\circ$ \citep{Sanchis-Ojeda2013}, although some other young systems show less direct evidence of misalignments \citep[NGTS-33b;][]{Alves2025}. The lack of younger misaligned systems favors slow/secular processes to explain misalignments seen in the mature planets \citep{Fabrycky2007, Albrecht2022}. However, the current sample size is too small and inhomogeneous to make definitive statements. The SOYSAUCE survey was designed to bridge this gap.

Multi-planet systems provide additional leverage in understanding these evolutionary pathways, particularly those in or near mean-motion resonances (MMR). Young planets are frequently found in resonant configurations and often exhibit high rates of transit timing variations \citep[TTVs;][]{Dai2024, LopezMurillo2026}. These systems are especially useful because they offer a potential pathway to measuring planetary masses in young systems \citep{Livingston2026}, where high levels of stellar activity often make traditional radial velocity mass determinations extremely challenging \citep{Blunt2023}.

In this paper, we present the validation, follow-up, and Rossiter-McLaughlin analysis of TIC~150070085\,b, as well as the discovery of a candidate signal TIC~150070085\,c. The inner planet was first identified as a candidate in \citet{Vach2024_occ}, but not validated at that time. In Section~\ref{sec:obs}, we describe our photometric and spectroscopic follow-up, including the Rossiter-McLaughlin (RM) data. Section~\ref{sec:transit_search} details our transit identification process, including the recovery of the 10.47-day primary planet and the detection of a second planet candidate. Following a derivation of stellar parameters in Section~\ref{sec:stellar_prop}, we provide a comprehensive analysis of the parent cluster, Alessi~84, and its $135\pm10$\,Myr age in Section \ref{sec:age}. Section~\ref{sec:planet} presents our transit analysis, while Section~\ref{sec:fp} covers our false-positive analysis. Finally, we discuss the implications of this system for planet evolution in Section~\ref{sec:discussion}.

\section{Observations} \label{sec:obs}

\subsection{\tess}

\starname\ was first observed by \tess\ in Sector 20 from 2019 December 25 to 2020 January 20, and re-observed in Sector 47 from 2021 December 31 to 2022 January 27. The target was pre-selected for 2-minute short cadence observations in Sector 47, but was only observed in \tess\ full frame images (FFIs) for Sector 20. The \tess\ data used in this analysis can be found in MAST \citep{tesslc_spoc,tesslc_ffi}. 

We extracted our \tess\ light curve following \cite{Barber2024_iras} and \cite{Vanderburg2019}, which includes a more robust handling of systematics removal in the presence of stellar variability. This approach has worked well on prior young planetary systems with \tess\ \citep[e.g.,][]{THYMEIII, Thao2024_1224}. We used the light curve with the fastest cadence available for each sector.

\subsection{LCOGT -- Photometry}

The \tess\ Follow-up Observing Program \citep[TFOP;][]{Collins2019_tfop}\footnote{\url{https://tess.mit.edu/followup}} obtained three full and three partial predicted transits of \planetname\ using the Las Cumbres Observatory Global Telescope \citep[LCO;][]{Brown2013_lcogt}. The full transits were from 2024 February 1, 2024 December 11, and 2025 January 1, all taken from McDonald Observatory in Fort Davis, Texas. Two partial transits taken with McDonald Observatory were from 2024 November 20 and 2025 December 23, and the last one was from 2024 October 20 taken using the Teide Observatory on the island of Tenerife. The team took all observations using the 1m telescopes and SINISTRO cameras with a $z_s$ filter and 43s exposures. 

LCO automated calibrated images using the standard LCOGT \texttt{BANZAI} pipeline \citep{McCully18}. We extracted differential photometry using \texttt{AstroImageJ} \citep{Collins17} with a set of hand-selected comparison stars. For the 2024 February, October and November and 2025 January observations, we used a 4.3" (11 pixel) radius circular aperture. For the 2024 December observation, we used a 5.8" (15 pixel) radius circular aperture, and for the 2025 December observation, we used a 3.9" (10 pixel) radius circular aperture. 

On 2026 March 6, we observed a transit of \planetname\ from the 0.4m telescope at Teide Observatory. We used the QHY600 camera with the Astrodon-Exo filter and 40s exposures. The QHY600 camera has a pixel scale of 0.74"/pixel.

As above, the LCO automatically performs basic reduction with the \texttt{BANZAI} pipeline. However, we extracted photometry for these using \texttt{Photutils} aperture photometry. A quick cross-check suggests using \texttt{AstroImageJ} would yield a consistent light curve. We used a 12 pixel (8.9") radius circular aperture. We tested combinations of 3--15 comparison stars, adopting the extraction that minimized the point-to-point scatter. We obtained a second, simultaneous observation of the transit using the 1m telescope at Teide Observatory in the $r'$ filter. However, we chose to exclude the observations due to systematics in the out-of-transit data.

In all observations, we detected the transit in the photometric aperture, confirming the transit is on-source down to the aperture size ($<4\arcsec$).

\subsection{LCOGT -- NRES}

We obtained six high-resolution ($R \simeq 53,000$) spectra of \starname{} using the Network of Robotic Echelle Spectrographs (NRES; \cite{Siverd2018}). These observations were conducted across multiple sites within the Las Cumbres Observatory \citep[LCO;][]{Brown2013_lcogt} global network. Spectra were taken between 2025 November 22 and 2025 December 2 with exposure times between 3600 and 5143 seconds per spectrum. At a reference wavelength of 5180 \AA, the spectra achieved a signal-to-noise ratio (SNR) ranging from approximately 12 to 45 per resolution element. 

We utilized the radial velocities (RVs) and associated uncertainties derived from the standard BANZAI-NRES reduction pipeline \citep{McCully2022}. The reduction process included a cross-correlation function (CCF) analysis against a best-fit Phoenix template to determine the barycentric velocities. 

The measured RVs for the system fluctuated between approximately 11.0 and 13.4\,km\,s$^{-1}$ during the observation window, with RV errors between 0.25 and 0.37\,km\,s$^{-1}$. Given the expected stellar jitter, this is consistent with no variation.

\subsection{TRES}

We obtained five spectra of \starname{} with the Tillinghast Reflector Echelle Spectrograph \citep[TRES;][]{tres} between 2024 April and 2025 February. TRES has a spectral resolution of R$\simeq$44,000 and is located on the Fred Lawrence Whipple Observatory (FLWO) 1.5 m Tillinghast Reflector telescope on Mount Hopkins, Arizona. The data were reduced and the radial velocities were extracted using the pipeline described in \cite{buchhave2010}.

\subsection{Speckle Imaging}

To check for companions or background stars below the resolution limits of \gaia{}, we obtained high-resolution speckle images of \starname{} using the `Alopeke instrument on the Gemini North 8\,m telescope in Maunakea, Hawaii. `Alopeke provided simultaneous data in two bands (562 nm and 832 nm). Images were processed following \citet{Howell2011}. We find no evidence of companions with magnitude differences $\Delta m_{562}<4$ and $\Delta m_{832}<5$ down to $\simeq0.1$\arcsec. The contrast curves are available on ExoFOP \citep{EXOFOP}.

\subsection{MAROON-X}
We used \maroonx\ to observe a transit of \planetname\ on UT 2025 December 23. \maroonx\ is a high-resolution (R$\simeq$85000) optical fiber-fed echelle spectrograph that splits light into red (649--920nm) and blue (500--663nm) channels \citep{Seifahrt2018, Seifahrt2020}. Due to variable clouds, the observations are a mix of 20-minute and 10-minute cadence. In total, we observed \starname\ for 7.9 hours, including pre- and post- transit baseline.

The relative radial velocities were extracted from the red and blue channels separately using the custom data extraction pipeline from the \maroonx\ team and the \texttt{SERVAL} pipeline \citep{Zechmeister2018}. \texttt{SERVAL} creates a stellar template by stacking the spectra and comparing an RV-shifted template to each observation. 

Several orders poorly converged during the \texttt{SERVAL} reduction, which were initially excluded from our analysis. For those on the blue channel, including these orders made no significant difference in the order-averaged RVs or the resulting RM analysis, so we opted to leave them out. For the red channel, the missing orders meant the precision was too low to detect the RM signal, particularly when we tested masking out orders with emission lines, so we opted to include these.

\citet{Silva2025} noted that telluric features can introduce systematic biases in RVs derived using \texttt{SERVAL} and similar pipelines. While the default pipeline masks tellurics, we may be more impacted by this issue because our observations span only a single night, meaning any template we build would have similar telluric features (data taken over long periods can mitigate this because the tellurics vary). Consistent with this, we observed that orders on the red channel with significant H$_2$O or O$_2$ telluric lines tended to show an overall negative slope. These seem to preserve the RM signal, but create a bias in the out-of-transit shape. We opted to remove RVs from seven orders in the red channel that showed such evidence of telluric contamination.

The changes to the red channel ultimately had a small impact on the analysis (see Section~\ref{sec:planet}), in large part because the uncertainties from the blue orders were much smaller. Indeed, even when we exclude the whole red channel we get a consistent $\lambda$ (1-2$\sigma$ from 0), albeit with larger uncertainties.

\section{Planet Detection}\label{sec:transit_search}

\starname\ was originally identified as hosting a candidate planet in \cite{Vach2024_occ}. As part of our validation process, we re-searched the system with a different transit search method to ensure we recovered the planet and to see if we could recover any additional transit-like signals.  

Following \cite{Barber2024_iras}, we used the updated \texttt{Notch \& LOCoR} search algorithm \citep[\texttt{N\&L};][]{Rizzuto2017}. We detrended the \tess\ light curve using a 0.5-day filtering window and calculated the change in the Bayesian Information Criterion (BIC) at each point for whether the inclusion of a trapezoidal transit model improved the light curve polynomial fit. We then used a box-least squares (BLS) on the BIC time series to identify periodic signals between 0.5 and 30 days, requiring a BLS SNR $>$8 (consistent with detection thresholds adopted in independent transit searches \citep{Fernandes2022,Vach2024_occ} and the \kepler\ and \tess\ missions to limit statistical false positives \citep{Jenkins2002,Jenkins2010,2021ApJS..254...39G}). We recovered the original signal identified in \cite{Vach2024_occ} (with a period of 10.47 days; \planetname) with a BLS SNR of 11. We additionally identified a signal with a period of 7.9 days with an SNR of 12. Further inspection of the light curve suggested this was likely a half-period alias. We explore the validity of this second candidate signal in Section \ref{sec:planetc}.

\section{Stellar Properties}\label{sec:stellar_prop}

\subsection{Spectral parameters from SPC}

We analyzed the TRES spectra using the \texttt{Stellar Parameter Classification} \citep[\texttt{SPC};][]{Buchhave2012} tool to derive stellar atmospheric parameters \teff, \logg, [m/H], and \vsini. In short, \texttt{SPC} cross correlates an observed spectrum against a grid of synthetic spectra based on Kurucz atmospheric models \citep{Kurucz1992}. This gave \teff=5999$\pm$50\,K, log(g)=4.42$\pm$0.10, [m/H]=0.00$\pm$0.08, and \vsini=20.1$\pm$0.5 km/s.

The \texttt{SPC} cross correlation does not account for macroturbulance when computing \vsini. Using a Least Squares Deconvolution method, we find \vsini\ $=18.2\pm0.4$\,\kms. This is in rough agreement with the value from \texttt{SPC} (to $\sim$3$\sigma$). We adopt a \vsini\ of $19\pm1$\,\kms\ for the remainder of this analysis.

\subsection{\teff, $L_{*}$ and $R_{*}$ from the spectral-energy distribution}

To determine $L_*$ and $R_*$, as well as provide an additional constraint on \teff, we fit the SED following \citet{Mann2016b} and using photometry from \citet{Skrutskie2006}, \citet{Henden2012}, \citet{allwise}, or \citet{Evans2018}. We compared the observed photometry to a grid of flux-calibrated templates from \citet{Heap2007}, \citet{Rayner2009}, and \citet{Falcon2011}, which have been supplemented with optical or NIR flux-calibrated spectra where available \citep{Mann2013c,Gaidos2014}. The templates spanned 0.36--2.4\um, but we used PHOENIX BT-SETTL atmosphere models \citep{Allard2013} to fill from 2.4--20\um. 

To compute \fbol, we integrated the resulting absolutely-calibrated spectrum, and derived $L_*$ by combining \fbol\ with the \gaia\ DR3 parallax. We estimated \teff\ from the BT-SETTL model fit (a free parameter of the fit) and $R_*$ from the Stefan-Boltzmann relation as well as from the scale factor between the model and the absolutely-calibrated spectrum \citep[the infrared flux method;][]{Blackwell1977}. 

\begin{figure}
	\begin{center}
		\includegraphics[width = 0.48\textwidth]{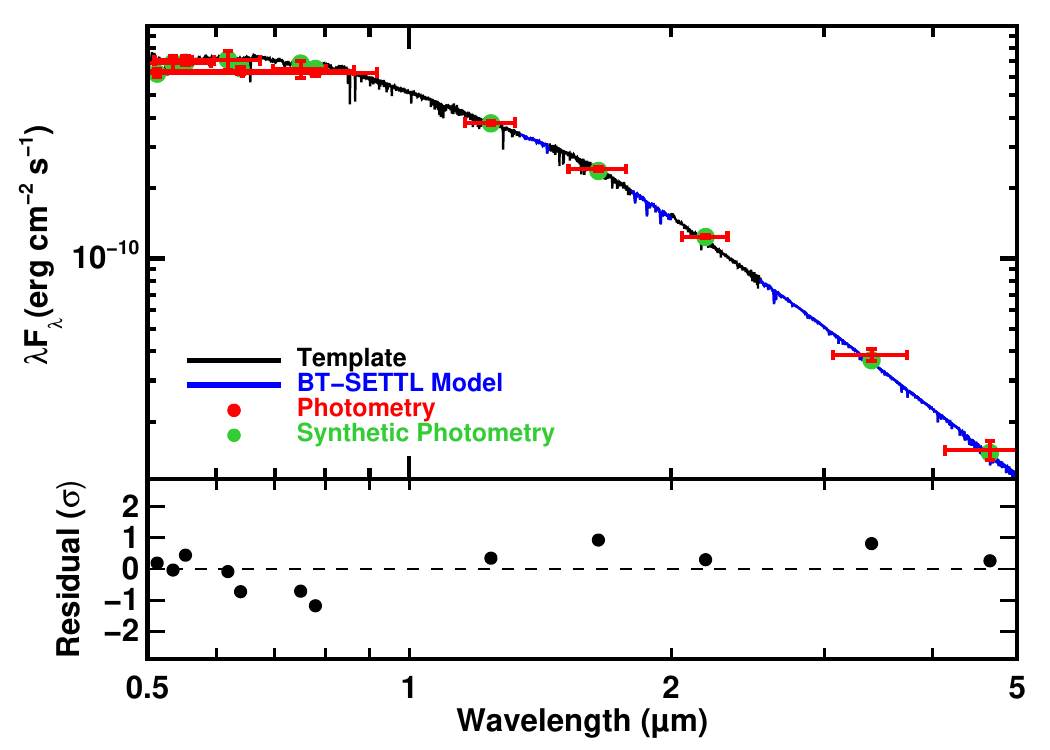}
		\caption{Representative result from our SED fitting procedure, including observed (red points) and synthetic (green) photometry, best-fit template (black line) and BT-SETTL model (blue) for \starname. The vertical error bars indicate the uncertainties in the photometry, while the horizontal bars indicate the filter width. The bottom panel shows the residual in units of standard deviations.\label{fig:sed}
        }
    \end{center}
\end{figure}

The resulting parameters were \fbol$=(1.05\pm0.06)\times10^{-9}$erg\,cm$^{-2}\,s^{-1}$, $L_*=1.37\pm0.08L_\odot$, \teff=$6053\pm70$\,K and $R_*=1.086\pm0.045R_\odot$. We found extinction to be low ($A_V<0.15$). We show an example fit in Figure~\ref{fig:sed}. The \teff\ was in excellent agreement with the SPC-based value, and using the SPC \teff\ with our $L_*$ gave us a consistent $R_*$.

\subsection{Model parameters from \texttt{StelPar}}

To derive $M_*$, we used the stellar parameter estimation and analysis tool \citep[\texttt{StelPar};][]{Fields2025}. \texttt{StelPar} compares observed photometry and the \gaia\ DR3 parallax to predictions from the PARSECv2.0 \citep{Nguyen2022_parsec} and DSEP-magnetic \citep{Feiden2012b,Feiden2013} model grids. The code was optimized and tested primarily on young stars, and reproduces observed stellar densities from transits of $\lesssim100$\,Myr stars.

We applied a Gaussian age prior from our group determination (Section~\ref{sec:age}). We initially used a uniform prior on \teff\ and $A_V$, but the degeneracy between the two yielded relatively poor overall constraints. We opted for a Gaussian prior on \teff\ from SPC results above. 

The final mass from \texttt{StelPar} was $1.16\pm0.06M_\odot$, with a radius of $R_*=1.15\pm0.06R_\odot$, consistent with all determinations above. We adopt this mass, but retain the SED-based $R_*$ for the rest of the analysis.

\subsection{Stellar inclination}
Following \cite{Masuda2020} and using the associated code from \cite{Fields2025}\footnote{\url{https://github.com/mjfields/cosi}}, we estimate the stellar inclination ($i_*$) from the \vsini\ measurements. Taking into account the rotation period (see Section \ref{sec:gyro}), stellar radius, \vsini, and associated uncertainties for each, we find $i_* > 69\degree$ at 1$\sigma$ ($>64\degree$ at 2$\sigma$). This is consistent with edge-on stellar rotation.

We report all final stellar parameters in Table \ref{tab:stellarparams}.

\begin{table}
    \centering
    \caption{Stellar parameters of \starname}
    \begin{tabular}{lcc}
    \hline
    \hline
    Parameter & Value & Source\\
    \hline
    \multicolumn{3}{c}{Identifiers}\\
    \hline
    TIC & 150070085 & \tess \\
    Gaia & 986757333119241216 & \gaia\ DR3 \\
    \hline
    \multicolumn{3}{c}{Astrometry}\\
    \hline
    $\alpha$ (deg) & 110.143200 & \gaia\ DR3 \\
    $\delta$ (deg) & 53.546980 & \gaia\ DR3 \\
    $\mu_\alpha$ (mas yr$^{-1}$) & $-2.542\pm0.012$ & \gaia\ DR3 \\
    $\mu_\delta$ (mas yr$^{-1}$) & $-30.3556\pm0.0091$ & \gaia\ DR3 \\
    $\pi$ (mas)  & $4.896\pm0.017$ & \gaia\ DR3 \\
    \hline
    \multicolumn{3}{c}{Photometry}\\
    \hline
    \tess\ (mag) & $10.5795\pm0.0066$ & \tess \\
    $G_\gaia$ & $10.9947\pm0.0028$ & \gaia\ DR3 \\
    $BP_\gaia$ & $11.3010\pm0.0034$ & \gaia\ DR3 \\
    $RP_\gaia$ & $10.5201\pm0.0040$ & \gaia\ DR3 \\
    \hline
    \multicolumn{3}{c}{Physical Properties}\\
    \hline
    \vsini\ (\kms) & $19\pm1$ & This work \\
    $i_*$ ($\degree$) & $>$69 & This work \\
    $\psi$ ($\degree$) & $23\pm11$ & This work \\
    $P_{rot}$ (days) & $2.886\pm0.036$ & \cite{Boyle2026_tars} \\
    \fbol\,(erg\,cm$^{-2}$\,s$^{-1}$) & $(1.05$$\pm$$0.06)$$\times$$10^{-9}$ & This work \\
    \teff\ (K) & $5999\pm50$ & This work \\
    $M_*$ ($M_\odot$) & $1.16\pm0.06$ & This work \\
    $R_*$ ($R_\odot$) & $1.086\pm0.045$ & This work \\
    $\rho_*$ ($\rho_\odot$) & $0.91\pm0.12$ & This work \\
    $L_*$ ($L_\odot$) & $1.37\pm0.08$ & This work \\
    Age (Myr) & $135\pm10$ & This work \\
    \hline
    \end{tabular}
    \label{tab:stellarparams}
\end{table}

\section{Parent population and age determination}\label{sec:age}

We can confirm \starname\ is young from its spectroscopic and photometric properties. However, age can be much more precisely determined from its parent population, which we investigate below.

\subsection{Selection of stars co-moving with \starname}
 
To identify candidate co-moving stars, we used the \texttt{FriendFinder}\footnote{\url{https://github.com/adamkraus/comove}},\footnote{alternatively referred to as \texttt{FindFriends} and \texttt{Comove}} algorithm \citep{THYMEV}, which leverages Gaia DR3 data to identify co-moving stellar populations. Because most stars do not have known radial velocities, the algorithm projects the six-dimensional coordinates of each target star into a five-dimensional space by predicting their proper motions and positions assuming a match to \starname, and then compares these predictions to \gaia’s three-dimensional map. Our primary goal was to identify enough comoving stars to calculate a precise age of the planet, so we prefer to identify a ``clean" list of members versus a complete one. We used tight search bounds (tangential velocity $<2$\kms, 3D distance $<25$ pc), which yielded 134 candidate comoving stars.

\starname\ has been previously clustered into a range of low-mass associations; OSCN 304 \citep[67 members;][]{Qin2023}, UPK 343 \citep[47 members;][]{Sim2019}, CWNU 1128 \citep[45 members;][]{He2022}, Theia 214 \citep[279 members;][]{Kounkel2019}, and Alessi 84 \citep[128 members;][]{Hunt2024}. These lists have significant overlap as they are identifying the same underlying structure but with varying methods. Combining with our \texttt{FriendFinder} list, we identify 334 unique comoving candidates across the six lists.

\begin{figure}
	\begin{center}
		\includegraphics[width = 0.48\textwidth]{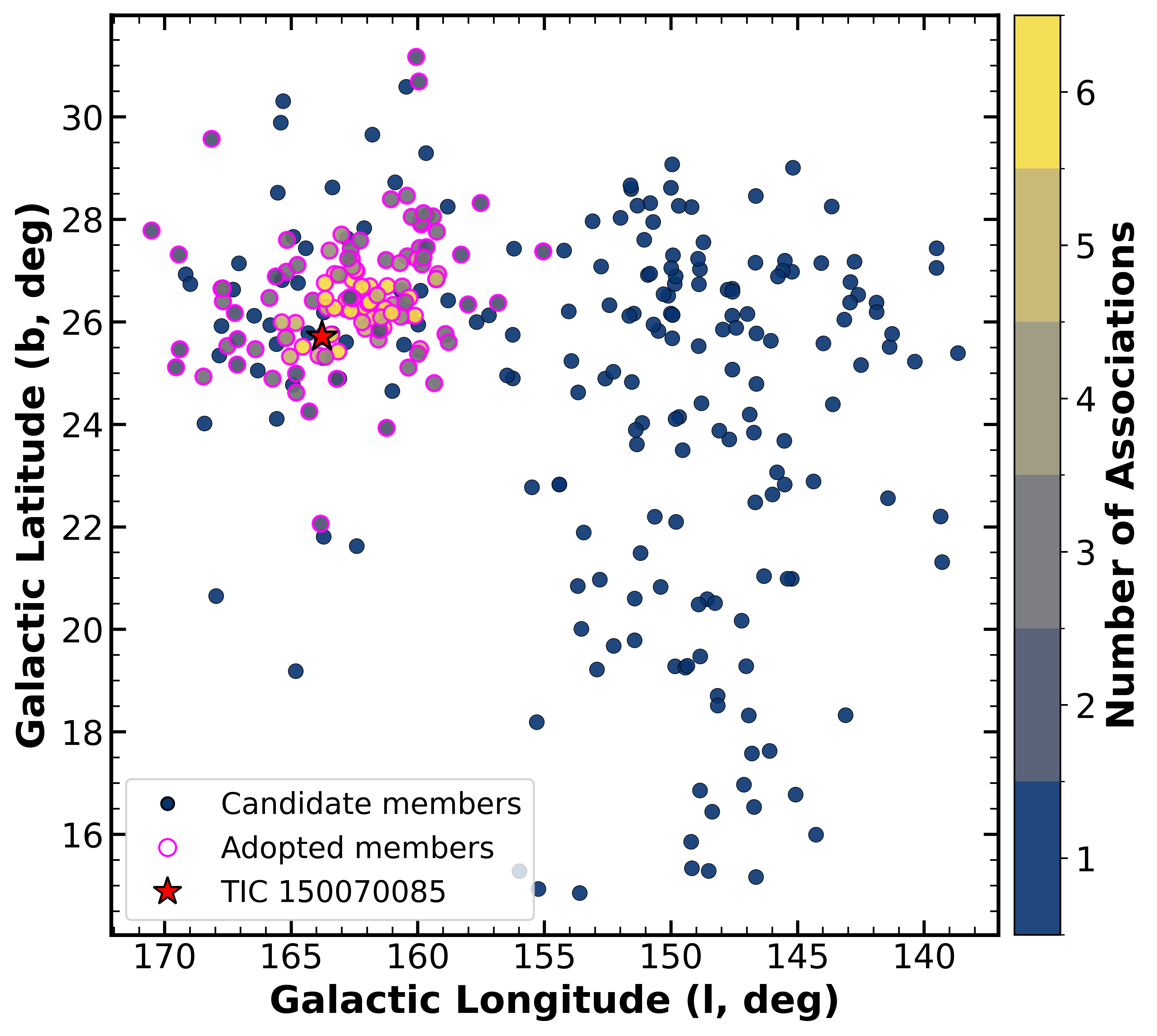}
		\caption{Adopted members of Alessi~84 (pink outlines) compared to all comoving candidates (circles) colored by the number of membership lists (associations) the star was identified a member in. \starname\ is shown as the red star. Requiring the candidate be in at least two of the membership lists primarily cuts members of Theia 214 that sit $\gtrsim5\degree$ from the remaining cluster. \label{fig:cluster}
        }
    \end{center}
\end{figure}

We keep all co-moving candidates that appear in at least two of the six membership lists (counting the \texttt{FriendFinder} list). This primarily removes stars from Theia 214 that sit $\gtrsim$5--10 deg from the remaining cluster (see Figure~\ref{fig:cluster}). These may be associated with the same group, but are not required for age dating the planet (and may create bias if the population is unassociated). We additionally cut three stars which disagree with the remaining rotation period sequence (see Section \ref{sec:gyro}). This cuts our final population to 124 stars. We list our final membership list in Table \ref{tab:group}. Because of the high proportion of stars from Alessi 84 in our final list (119 of the 124), we refer to the cluster as Alessi 84 for the remainder of this work. 

\subsection{Age Analysis} \label{age analysis}

The associations that \starname\ has been clustered into range in age from 97--158 Myr \citep{Kounkel2019, Kounkel2020, Qin2023, Hunt2023}. Although these estimates agree with one another within uncertainties, this is primarily because many of the uncertainties are large (as high as 90\,Myr). Prior studies have achieved far higher precision on similarly low-mass groups \citep[e.g.,][]{THYMEVII, Wood2023, Thao2024_1224}. Since the literature age determinations are reliant almost entirely on properties of the population's CMD, we should be able to improve the age by combining multiple aging metrics \citep[e.g.,][]{Barber2025_toi2076e}.

\subsubsection{Isochronal/CMD Modeling}

We compared the \gaia\ photometry to model isochrones using a mixture model as described in the appendix of \cite{THYMEVI}. The mixture model considers two populations; the first is a single-star single-age sequence derived from the model isochrone, and the second is an outlier population. The outliers are anything that causes the star to deviate from the isochrone, including non-members, binaries, or stars with poor parallaxes or photometry.

We model the main population with two free parameters: the age ($\tau$) and the overall extinction ($E(B-V)$). We model the outlier population as an offset from the primary population ($Y_B$), a variance ($V_B$) around that offset, and a population amplitude ($P_B$). The final parameter, $f$, accounts for underestimated uncertainties in the input parameters or additional sources of scatter (e.g., variation in extinction between sources). We assumed Solar metallicity as has been found for most young associations near the Sun \citep{Spina2017}, although we tested slightly super-Solar ([M/H]=$+0.2$) and sub-Solar ([M/H]=$-0.2$) grids as a check. 

\begin{figure}
    \centering
    \includegraphics[width=.48\textwidth]{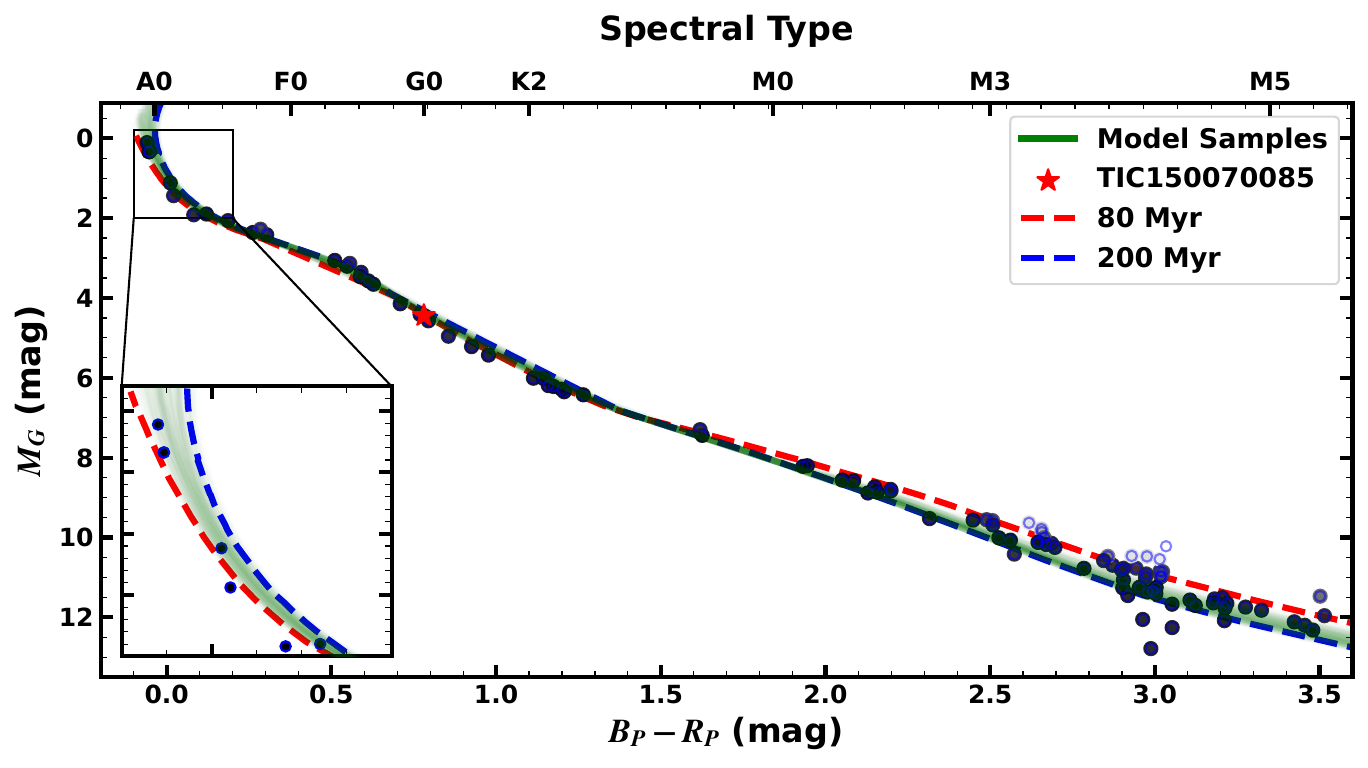}
    \caption{\gaia\ color-magnitude diagram of candidate members of \assoc. Points are color-coded by their outlier likelihood (darker indicates more likely part of the single-age single-star population). Green lines show 100 random draws from the MCMC posterior. Dashed lines indicate 80\,Myr and 200\,Myr isochrones from PARSEC for reference. The inset around the high-mass stars highlights how the handful of high-mass stars disfavor ages above $\simeq$150\,Myr while the pre-MS late-type stars strongly disfavor ages below $\simeq100$\,Myr. }
    \label{fig:cmd}
\end{figure}

We used isochrones from PARSECv2.0 \citep{Nguyen2022_parsec} and the Dartmouth stellar evolution program \citep[DSEP;][]{Dotter2008} with magnetic-enhancement \citep[DSEP-mag;][]{Feiden2016}. PARSEC favored a much lower age ($128^{+10}_{-8}$\,Myr) than DSEP-mag ($171^{+14}_{-22}$\,Myr), although consistent at 1.8$\sigma$. DSEP-mag models do not include B/A/F stars, which are challenging to fit with ages above 150\,Myr (Figure~\ref{fig:cmd}). This is likely the reason for the larger uncertainty and higher age from DSEP models. The pre-main-sequence M dwarfs (and lack of pre-main-sequence early Ms) disfavor younger ages ($<100$\,Myr), but this is true for either grid. Because of the better coverage, we adopted the PARSEC results. 

Metallicity changes at the 0.2~dex level had a small impact ($5$\,Myr difference). Solar-metallicity provided the best fit, which we adopted. However, we expanded our uncertainties to account for metallicity effects, giving a final isochronal age of $128^{+12}_{-10}$\,Myr

\subsubsection{Gyrochronology}\label{sec:gyro}

We pulled rotation periods for the candidate members from the TARS catalog \citep{Boyle2026_tars}, which includes rotation periods for $>$1M stars in the Solar neighborhood, with uncertainties assigned following \cite{Boyle2025a}. From the original 127 cluster members, TARS contained rotation periods for 28 stars that passed all quality flags (which test for contaminating stars, binarity, and variations in detected rotation period across multiple sectors). This hit rate is similar to expectations; few to no mid-to-late M dwarfs will be bright enough for TARS at this distance, some stars are cut by the binarity flags, and a few will be missed by chance (TARS completeness is not 100\% even for fast-rotators) or not be members. 

\begin{figure}
    \centering
    \includegraphics[width=0.99\linewidth]{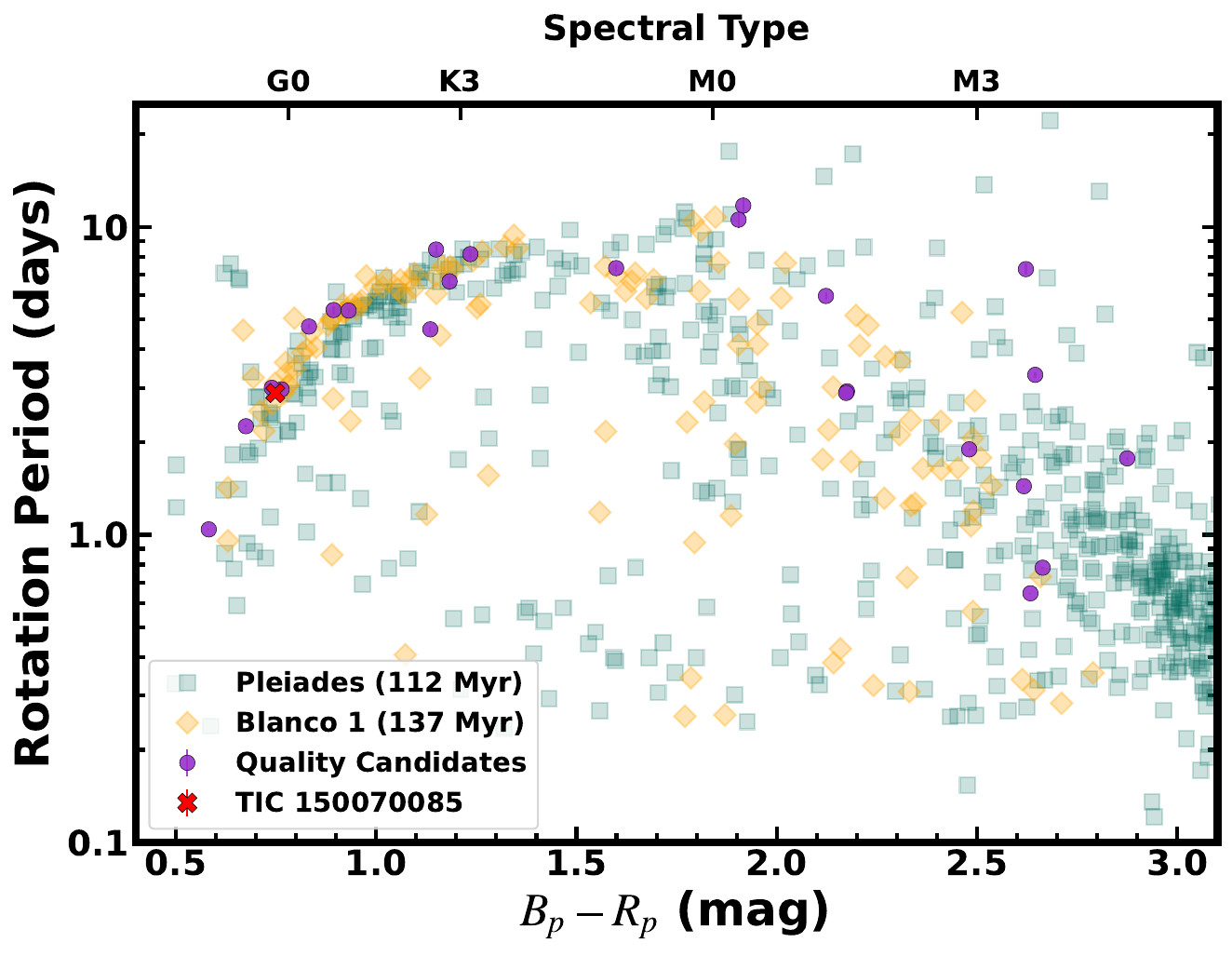}
    \caption{Rotation periods of cluster members (purple circles) comoving with \starname\ (red x). The rotation periods for the Pleiades \citep{Rebull2016} and Blanco 1 \citep{Gillen2020} (green squares and orange diamonds, respectively) are shown for reference.}
    \label{fig:prot}
\end{figure}

TARS finds the rotation period of \starname\ to be $2.886\pm0.036$ days, consistent with group membership (Figure \ref{fig:prot}). Three stars had rotation periods which disagreed with the remaining sequence. We chose to drop these stars from our membership list as likely nonmembers, leaving 25 stars with high reliability rotation periods.

Using the rotation periods, rotation period uncertainties, and effective temperatures in the TARS catalog, we calculated an age posterior for each star using \texttt{gyro-interp} \citep{Bouma2023_gyrointerp}. \texttt{gyro-interp} requires the effective temperature to be 3800--6200\,K, which cuts our sample to 14 stars for which we can calculate an age posterior. Following \cite{Bouma2023_gyrointerp}, we used \texttt{PosteriorStacker} \citep{Baronchelli2020_posteriorstacker} to combine the age posterior from each star and infer the group age in a hierarchical Bayesian framework, viable since \texttt{gyro-interp} adopts a uniform age prior. We combined the individual star estimates to get a group age of $164\pm35$ Myr.

\subsubsection{Variability}

\cite{Barber2023} mapped out the Skumanich-like correlation between excess uncertainties in \gaia\ photometry \citep{Riello2021} and stellar age. This takes advantage of variable stars having higher photometric uncertainties and the relationship between stellar variability and age. This method has proven useful for estimating approximate ages over a wide range of diffuse young populations where contamination is expected to be an issue \citep[e.g.,][]{Sun2023,Thao2024_1224,Distler2026}. We used \texttt{EVA}\footnote{\url{https://github.com/madysonb/EVA}} (Excess Variability-based Age) to calculate a variability age in \gaia$_G$, \gaia$_{B_P}$, and \gaia$_{R_P}$ bands. Following \cite{Barber2023}, we combined each age using a weighted mean and find a variability age of $178_{-50}^{+93}$ Myr. Due to the relatively small group size, the age uncertainty is quite large, as expected. Though it agrees with the other age estimates, \texttt{EVA} provides comparatively weak constraints.

\subsubsection{Combining age estimates}

\begin{figure}
    \centering
    \includegraphics[width=0.99\linewidth]{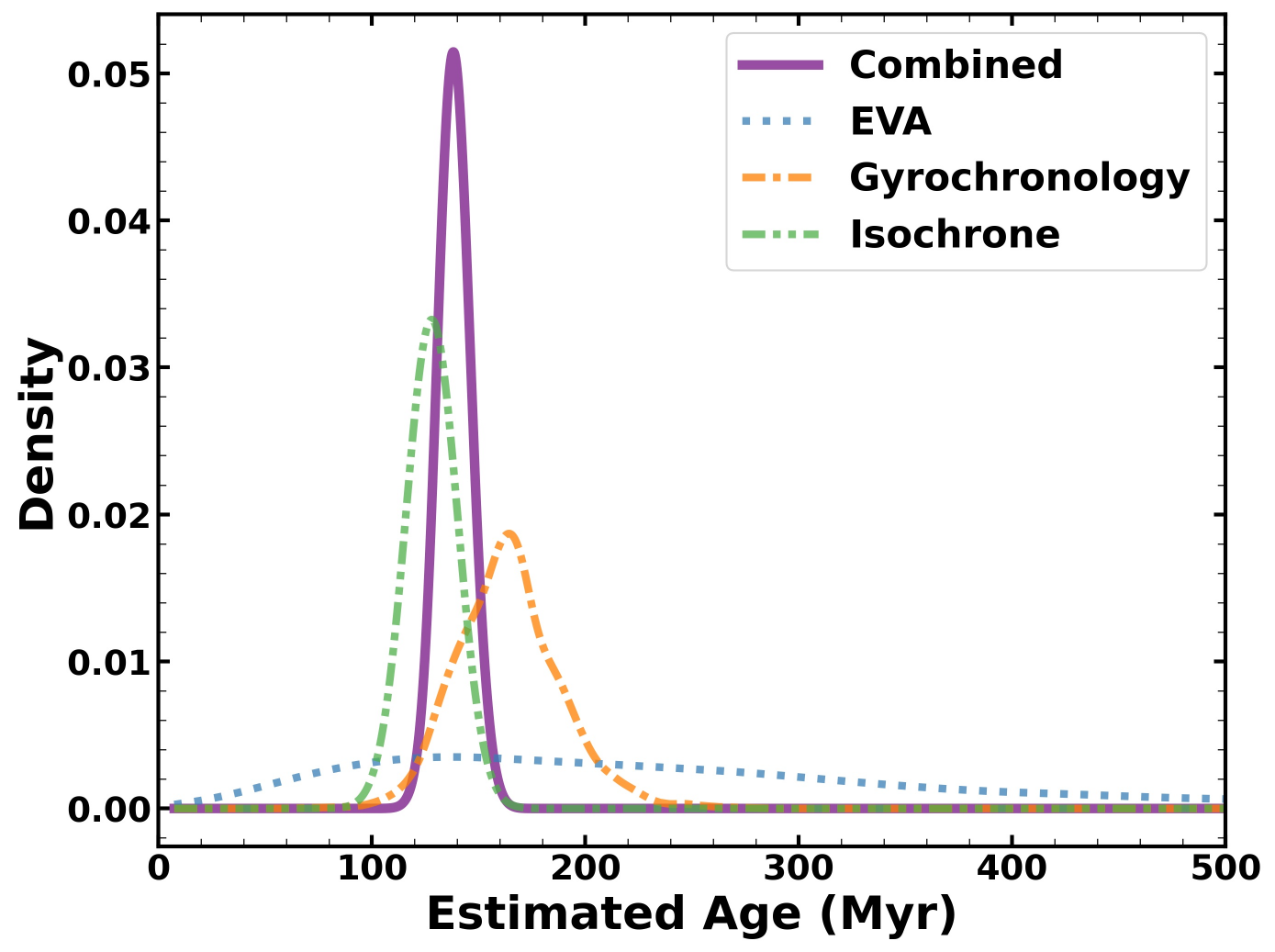}
    \caption{Individual age estimates from variability (EVA), gyrochronology, and the isochrone fit, as well as the combined final age of Alessi 84 ($135\pm10$ Myr; purple). All methods agree, although the isochrone fit is the most precise and dominates the final posterior. }
    \label{fig:combined}
\end{figure}

To determine a final age estimate for \starname, we combined the likelihood distribution from each individual age estimate, as shown in Figure \ref{fig:combined}. Using the isochronal modeling, gyrochronology, and variability ages, we find an overall age of $135\pm10$\,Myr for Alessi 84 and \starname. This is in agreement with previous age determinations but provides a much tighter constraint.

\subsubsection{Additional age checks}

We performed two additional age checks on \starname\ which we did not include in the combined final age estimate; lithium and \texttt{GaiaChrones} ages.

No stars in our final cluster membership list have an archival lithium measurement. However, we can use the \maroonx\ out-of-transit spectra to check for lithium in \starname. We estimated an equivalent width of $118\pm3$ m\AA. Using the empirical relation from \cite{Jeffries2023_eagles}, this suggests an age of $\sim$100 Myr. At this temperature and age range, lithium depletes slowly, leading to large uncertainties on the estimate (factor of $\simeq$2). However, the presence of lithium does confirm the star is young and therefore a likely member of Alessi 84.

We additionally ran \texttt{GaiaChrones} (Barber et al. in prep) on the group. \texttt{GaiaChrones} relates \gaia\ XP spectra and distances to stellar age using the \texttt{XGBoost} \citep{XGboost} machine learning algorithm trained on a set of known young clusters, with a mixture model to account for binaries and cluster interlopers. In essence, the method creates an empirical isochronal model. Using \texttt{GaiaChrones}, we find a consistent age of $142\pm12$\,Myr. Because the isochronal model and \texttt{GaiaChrones} probe the same underlying physics, we choose not to use both in our combined age, but this nicely matches our final adopted age (135$\pm$10\,Myr).

\section{Planet Properties} \label{sec:planet}

\subsection{\texttt{Juliet} Fit}\label{sec:juliet}

\begin{figure*}
    \centering
    \includegraphics[width=0.99\linewidth]{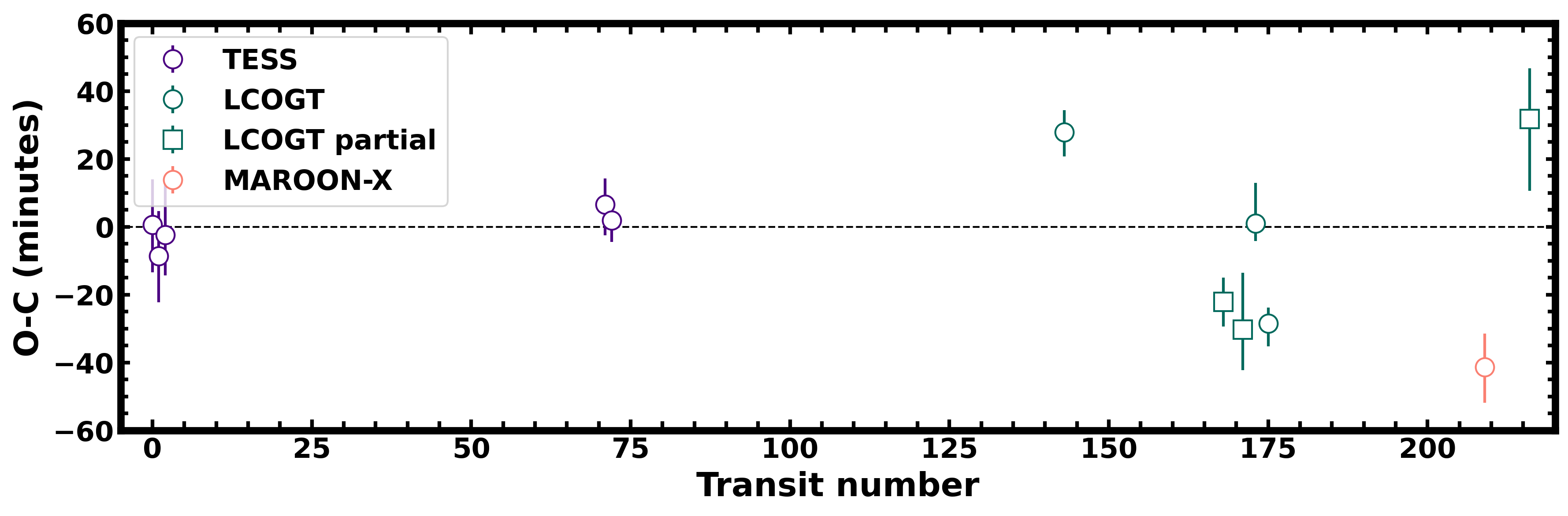}
    \caption{The observed minus expected transit timing of each photometric transit. The \tess\ data (purple) does not show an obvious offset, but the higher-cadence (43-second) ground-based photometry (green) and the \maroonx\ transit (pink) shows a clear TTV signal. The timings of the first three \tess\ transits are not well constrained due to the sparse, 30-minute data sampling.}
    \label{fig:ttv}
\end{figure*}

While the \tess\ data alone does not show significant transit timing variations (see Figure \ref{fig:ttv}), the additional ground-based data revealed a TTV signal with an amplitude of $\gtrsim$30 minutes. Because the Rossiter-McLaughlin fit is reliant on the transit parameters \citep[see][]{Hirano2020}, we opted to fit the photometric data using \texttt{Juliet} \citep{juliet}. \texttt{Juliet} enables us to simultaneously model the transit parameters, the TTV signal, and the stellar variability. We simultaneously fit the \tess\ data and ground-based data of full transits. We modeled the transit using \texttt{BATMAN} \citep{batman}, while we handled the stellar signal in the \tess{} data using the double simple harmonic oscillator (SHO) kernel Gaussian Process (GP) \citep[with the implementation in \texttt{celerite};][]{celerite}. There was not enough baseline to tune a GP on the ground-based data, so we use a simple linear trend line (which worked well for all transits). 

We opted to not include ground-based partial transit observations of \planetname\ in the \texttt{Juliet} analysis because they greatly increased the number of free parameters and provided no extra leverage on the transit parameters. Instead, we use the partial transits to further explore the TTV signal (see Section~\ref{sec:partials})

We fit for 33 parameters in total. For the transit fit, we fit all photometry with a common planet period ($P$), time of inferior conjunction ($T_c$), planet-to-star radius ratio ($R_P/R_*$), impact parameter ($b$), and stellar density ($\rho_*$). We assumed a circular orbit (locking eccentricity to 0). We fit for two quadratic limb-darkening parameters for each filter ($q_{1,TESS}$, $q_{2,TESS}$, $q_{1,z_s}$, $q_{2,z_s}$). For the GP, we fit for the GP period ($P$), the standard deviation of the process ($\sigma$), the quality factor ($Q$), the difference between the quality factors of the first and the second modes ($dQ$), and the fractional amplitude of the secondary mode compared to the primary ($f$). For each ground transit, we fit for a slope ($m$), with the linear model centered at the transit midpoint. Additionally, for each dataset, we fit for a constant flux offset ($c_{flux}$), and a jitter term ($\sigma_w$). Finally, we fit each transit with a unique transit midpoint. 

Most parameters were allowed to float with only physical limitations. We applied weak Gaussian priors to the individual transit midpoints (centered on their expected timing assuming a linear ephemeris). For the limb-darkening parameters, we used \texttt{LDTK} \citep{Parviainen2015_LDTK} and applied Gaussian priors that account for model differences. Since small planets and planets in compact multiplanet systems show low eccentric orbits \citep{Gilbert2025, Gilbert2026}, we locked eccentricity to 0 and placed a Gaussian prior on $\rho_*$ based on the $R_*$ and $M_*$ determined in Section \ref{sec:stellar_prop}. We widened the 1$\sigma$ prior to three times the expected uncertainty based on $R_*$ and $M_*$ to account for small deviations from a perfectly circular orbit. We list all priors in Table \ref{tab:transitFit}.

For the fit, we used \texttt{Juliet}'s implementation of Dynesty's importance nested sampling \citep{Speagle2020_dynesty, Koposov2025_dynesty} with 2000 live-points and dynamic sampling. 

\begin{figure*}
    \centering
    \includegraphics[width=0.4\linewidth]{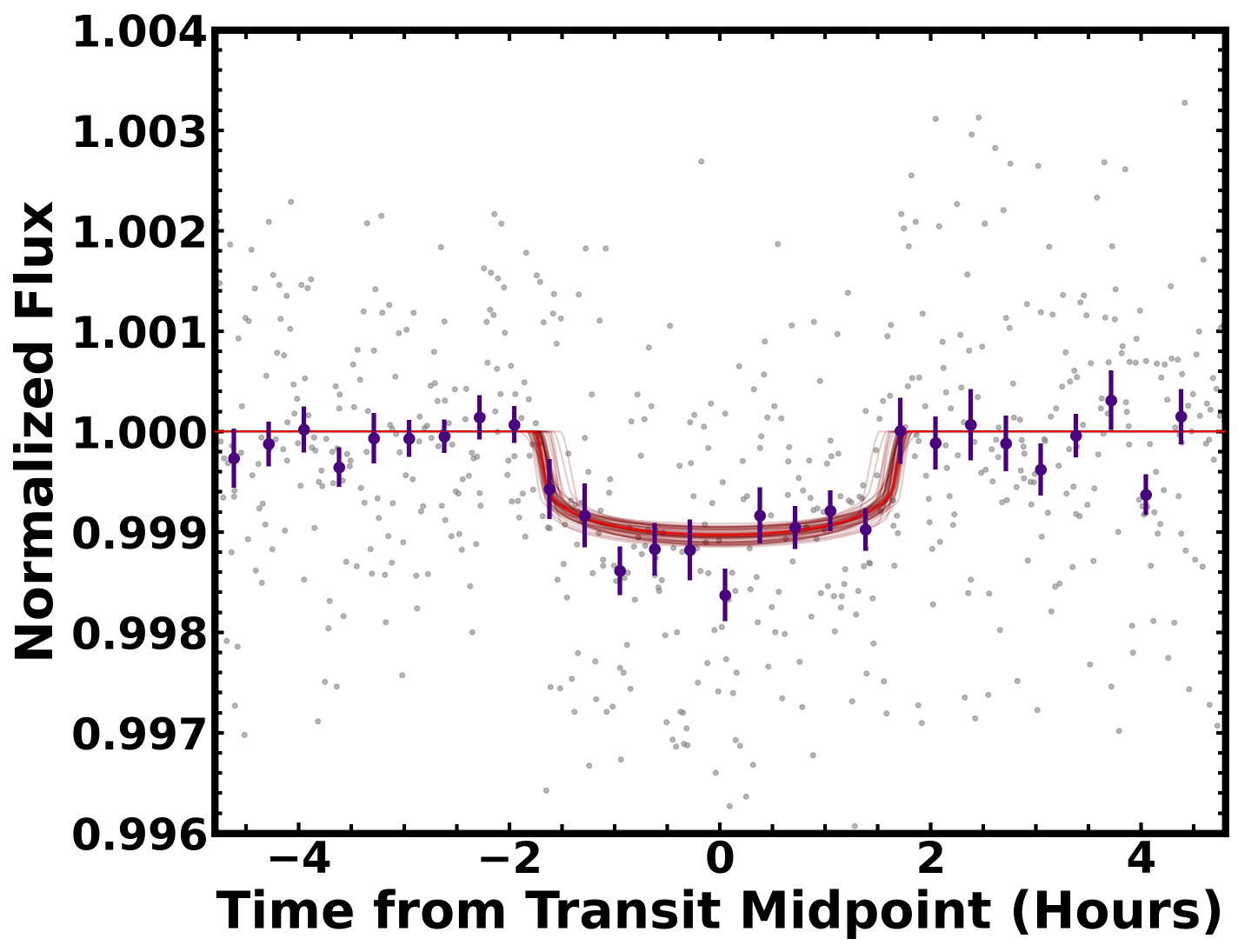}
    \includegraphics[width=0.4\linewidth]{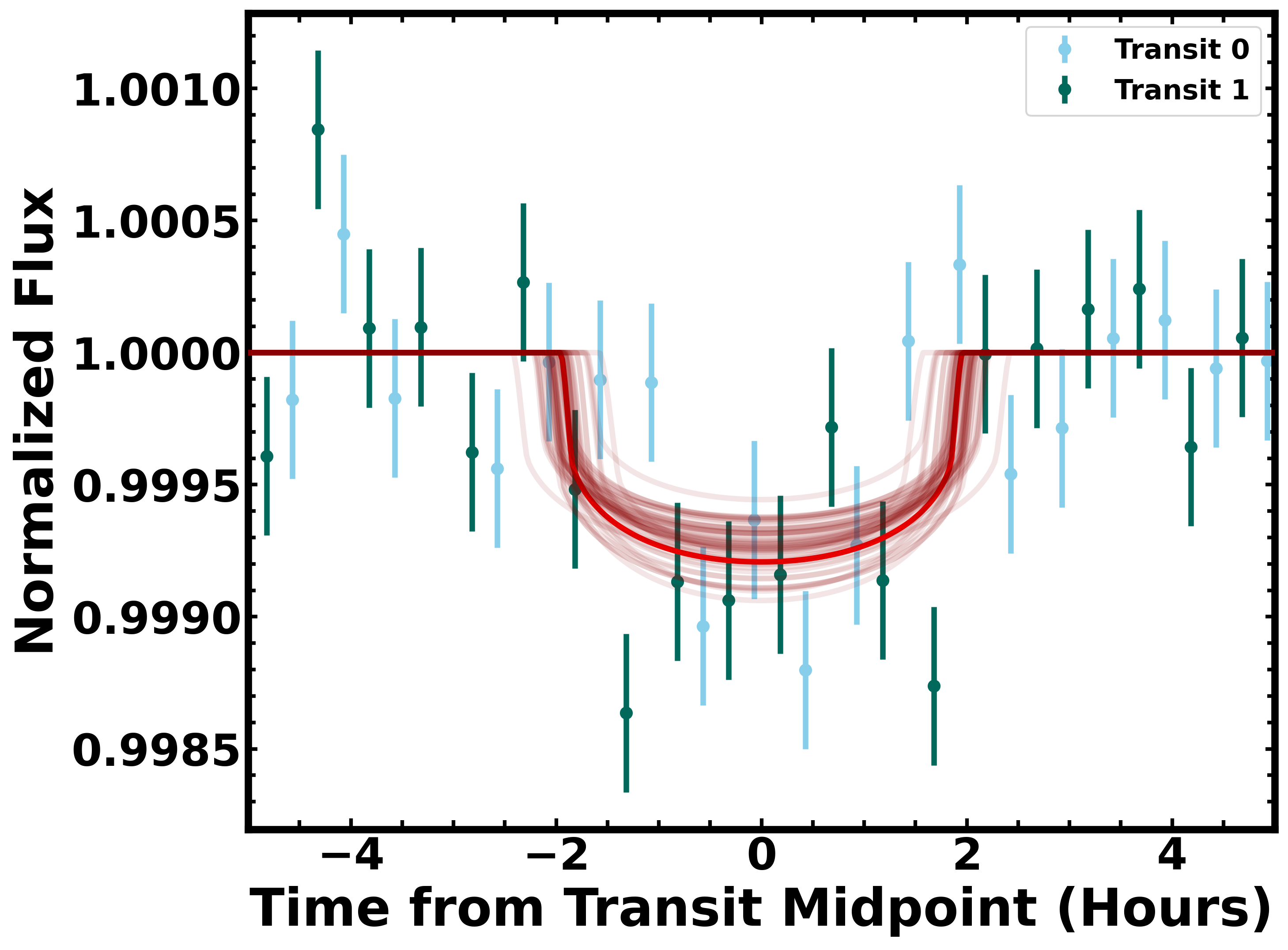}\\
    \includegraphics[width=0.9\linewidth]{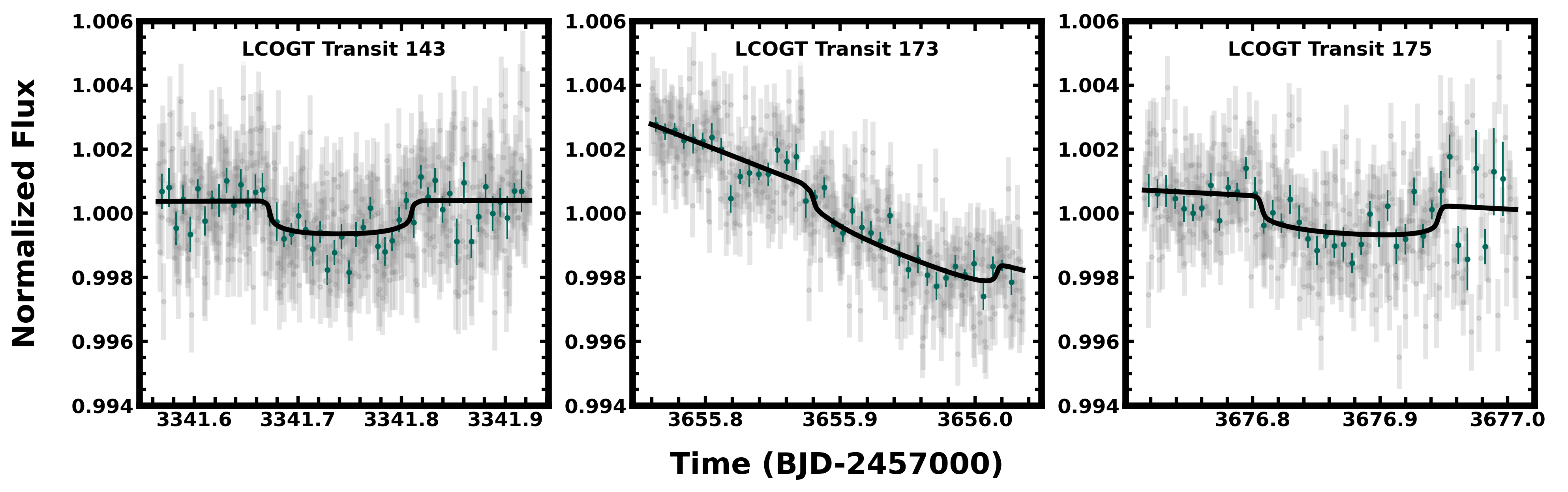}\\

    \caption{Top left: Phase-folded \tess\ light curve (gray), with stellar variability removed and corrected for transit timing offsets, showing the transit of \planetname. The data is binned to 20-minute intervals (purple) for clarity. The best-fit transit models are shown in bright red, with 50 randomly-drawn models from the posterior in dark red. Top right: Phase-folded \tess\ light curve showing the two transits of candidate \starname\ c (blue and green). The best-fit transit model is shown in bright red, with 50 randomly-drawn models from the posterior in dark red. Bottom: the individual ground-based transits of \planetname\ from LCOGT (gray) binned to 10-minute intervals (green) with the best-fit transit model (black). Each are labeled with the transit number from the first detected \tess\ transit (Transit 0).}
    \label{fig:phased}
\end{figure*}

\begin{figure*}
    \centering
    \includegraphics[width=0.9\linewidth]{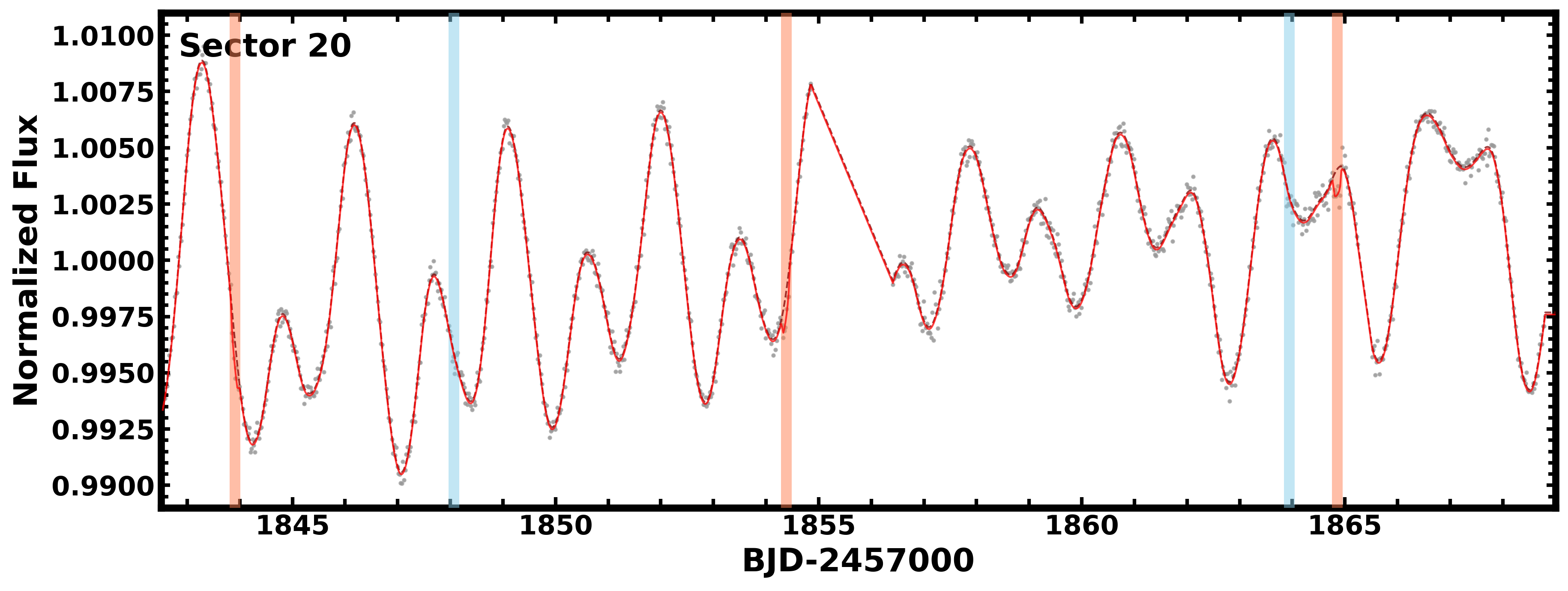}\\
    \includegraphics[width=0.9\linewidth]{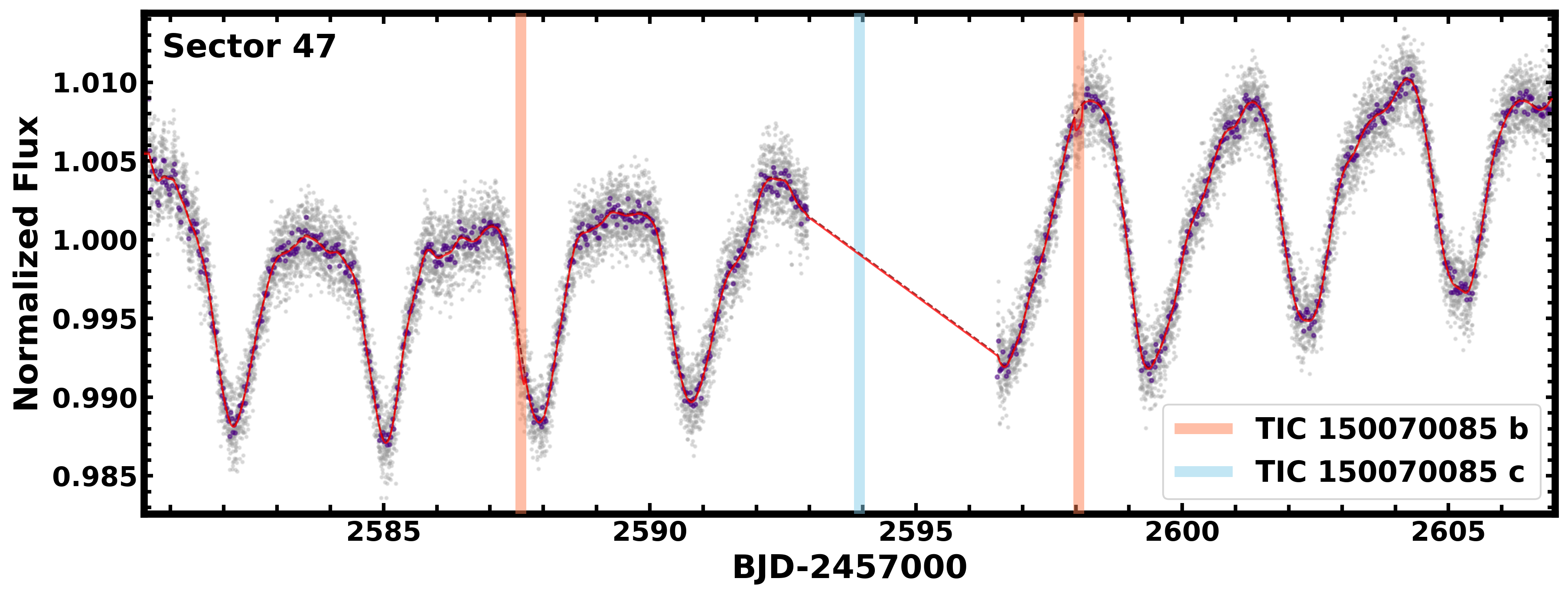}
    \caption{Sector 20 (top) and Sector 47 (bottom) of the \tess\ light curve (gray), binned to 30-minute intervals (purple), with the GP stellar variability model shown as the red line. The transit locations of \planetname\ are highlighted in pink, and the locations of candidate \starname\ c are highlighted in blue. Due to the data gap, \starname\ c was not observed in Sector 47, complicating validation of the signal.}
    \label{fig:gp}
\end{figure*}

\begin{table*}[]
    \centering
    \caption{Transit fit parameters and priors for \planetname}
    \begin{tabular}{lccc}
    \hline
    \hline
    Description & Parameter & Prior$^\alpha$ & Value\\
    \hline
    \multicolumn{4}{c}{Transit Parameters}\\
    \hline
    time of inferior conjunction & $T_c$ (BJD-2457000) & ... & $1843.9071^{+0.0058}_{-0.0057}$\\
    orbital period & $P$ (days) & ... &  $10.474230^{+0.000045}_{-0.000044}$\\
    planet-to-star radius ratio & $R_P/R_*$ & $U(0,1)$ & $0.0306^{+0.0013}_{-0.0012}$ \\
    impact parameter & $b$ & $U(0,1)$ & $0.53^{+0.15}_{-0.20}$\\
    stellar density & $\rho_*$ ($\rho_\odot$) & $N(0.91,0.35)$, $\rho_*>0$ & $1.07 \pm 0.34$\\
    \hline
    \multicolumn{4}{c}{Limb-darkening Coefficients}\\
    \hline
    linear \tess\ & $q_{1,\tess}$ & $N(0.376, 0.1)$ & $0.388\pm0.098$\\
    quadratic \tess\ & $q_{2,\tess}$ & $N(0.175, 0.05)$ & $0.179^{+0.049}_{-0.050}$\\
    linear $z_s$ & $q_{1,z_s}$ & $N(0.318, 0.1)$ & $0.344^{+0.097}_{-0.096}$\\
    quadratic $z_s$ & $q_{2,z_s}$ & $N(0.17, 0.05)$ & $0.177\pm0.049$\\
    \hline
    \multicolumn{4}{c}{Variability Parameters}\\
    \hline
    GP period & $P$ (days) & $N(2.88, 0.1)$ & $2.894^{+0.019}_{-0.017}$\\
    GP standard deviation & $\sigma$ & ... & $0.0173^{+0.0064}_{-0.0056}$\\
    GP quality factor & $Q$ & ... & $38.3^{+32.0}_{-19.7}$\\
    GP difference between modes & $dQ$ & ... & $0.069^{+1.229}_{-0.065}$\\
    GP fractional amplitude & $f$ & ... & $0.0106^{+0.0146}_{-0.0056}$\\
    flux offset \tess\ & $c_{flux,\tess}$ & ... & $0.00007\pm0.00037$\\
    flux offset $z_{s,1}$ & $c_{flux,z_{s,1}}$ & ... & $-0.000386^{+0.000066}_{-0.000064}$\\
    flux offset $z_{s,2}$ & $c_{flux,z_{s,2}}$ & ... & $-0.000521^{+0.000066}_{-0.000065}$\\
    flux offset $z_{s,3}$ & $c_{flux,z_{s,3}}$ & ... & $-0.000438^{+0.000068}_{-0.000069}$\\
    flux slope $z_{s,1}$ & $m_{flux,z_{s,1}}$ & ... & $0.00010^{+0.00058}_{-0.00059}$\\
    flux slope $z_{s,2}$ & $m_{flux,z_{s,2}}$ & ... & $-0.01645^{+0.00102}_{-0.00088}$\\
    flux slope $z_{s,3}$ & $m_{flux,z_{s,3}}$ & ... & $-0.00210^{+0.00079}_{-0.00076}$\\
    flux jitter \tess\ & $\sigma_{w,\tess}$ (ppm) & ... & $0.52^{+0.33}_{-0.35}$\\
    flux jitter $z_{s,1}$ & $\sigma_{w,z_{s,1}}$ (ppm) & ... & $0.49\pm0.34$\\
    flux jitter $z_{s,2}$ & $\sigma_{w,z_{s,2}}$ (ppm) & ... & $0.52^{+0.33}_{-0.35}$\\
    flux jitter $z_{s,3}$ & $\sigma_{w,z_{s,3}}$ (ppm) & ... & $0.51\pm0.34$\\
    \hline
    \multicolumn{4}{c}{Transit Midpoints}\\
    \hline
    Transit 0 & $T_0$ (BJD-2457000) & $N(1843.89, 0.1)$ & $1843.908^{+0.010}_{-0.011}$\\
    Transit 1 & $T_1$ (BJD-2457000) & $N(1854.36, 0.1)$ & $1854.376^{+0.010}_{-0.011}$\\
    Transit 2 & $T_2$ (BJD-2457000) & $N(1864.84, 0.1)$ & $1864.8533^{+0.0129}_{-0.0084}$\\
    Transit 71 & $T_{71}$ (BJD-2457000) & $N(2587.57, 0.1)$ & $2587.5822^{+0.0048}_{-0.0061}$\\
    Transit 72 & $T_{72}$ (BJD-2457000) & $N(2598.05, 0.1)$ & $2598.0530^{+0.0039}_{-0.0034}$\\
    Transit 143 & $T_{143}$ (BJD-2457000) & $N(3341.73, 0.05)$ & $3341.7414\pm0.0045$\\
    Transit 173 & $T_{173}$ (BJD-2457000) & $N(3655.95, 0.05)$ & $3655.9481^{+0.0127}_{-0.0043}$\\
    Transit 175 & $T_{175}$ (BJD-2457000) & $N(3676.88, 0.05)$ & $3676.8765^{+0.0039}_{-0.0022}$\\
    \hline
    \hline
    \multicolumn{4}{c}{Derived Parameters}\\
    \hline
    planet radius & $R_P$ ($R_\oplus$) & ... & $3.62^{+0.21}_{-0.20}$\\
    transit duration (first-to-fourth contact) & $t_{dur}$ (days) & ... & $0.1448^{+0.0033}_{-0.0074}$\\
    semi-major axis to stellar radius ratio & $a/R_*$ & ... & $20.58^{+1.99}_{-2.48}$\\
    inclination & $i(\degree)$ & ... & $88.51^{+0.63}_{-0.66}$\\
    equilibrium temperature$^{\dagger}$ & $T_{eq}$ (K)& ... & $945.06_{-42.89} ^{+62.73}$\\
    \hline
    \multicolumn{4}{l}{$\alpha$ $U(a,b)$ indicates a uniform distribution from $a$ to $b$.}\\
    \multicolumn{4}{l}{\,\,\,\,\,$N(a,b)$ indicates a normal distribution centered at $a$ with a standard deviation of $b$.}\\
    \multicolumn{4}{l}{$\dagger$ assuming zero albedo}
    \end{tabular}
    \label{tab:transitFit}
\end{table*}

The final fit parameters are given in Table~\ref{tab:transitFit}, and we show the overall light curve in Figure~\ref{fig:gp} and phased transits in Figure~\ref{fig:phased}. Overall, the GP did an excellent job handling the variability seen in the \tess{} data. The ground-based transits were reasonably explained with a simple linear fit, although all three show higher-order structure, likely due to atmospheric or instrumental systematics rather than missing stellar variability (based on the rapid timescale of the variations).

\subsection{Partial transit observations}\label{sec:partials}

We fit each partial transit with a unique transit midpoint ($T_X$) and stellar variability model, but lock the planet parameters to the median fit values from Section \ref{sec:juliet}. We model the transit using \texttt{BATMAN} and fit the out-of-transit variability using a third-order polynomial in time (with $T=0$ set to the first datapoint). The partials required a higher-order out-of-transit fit likely because of an inability to remove airmass-dependent or other Earth-based signals without a longer baseline. We applied weak Gaussian priors to $T_X$ centered at the expected timing assuming a linear ephemeris and a width of 0.05 days (72 minutes). These are much broader than the constraints from the data ($<10$\,minutes), but prevent the walkers from fitting random stellar or atmospheric variability as though it were a transit. For the limb-darkening, we used \texttt{LDTK} and applied Gaussian priors as above.

\begin{figure}
    \centering
    \includegraphics[width=\linewidth]{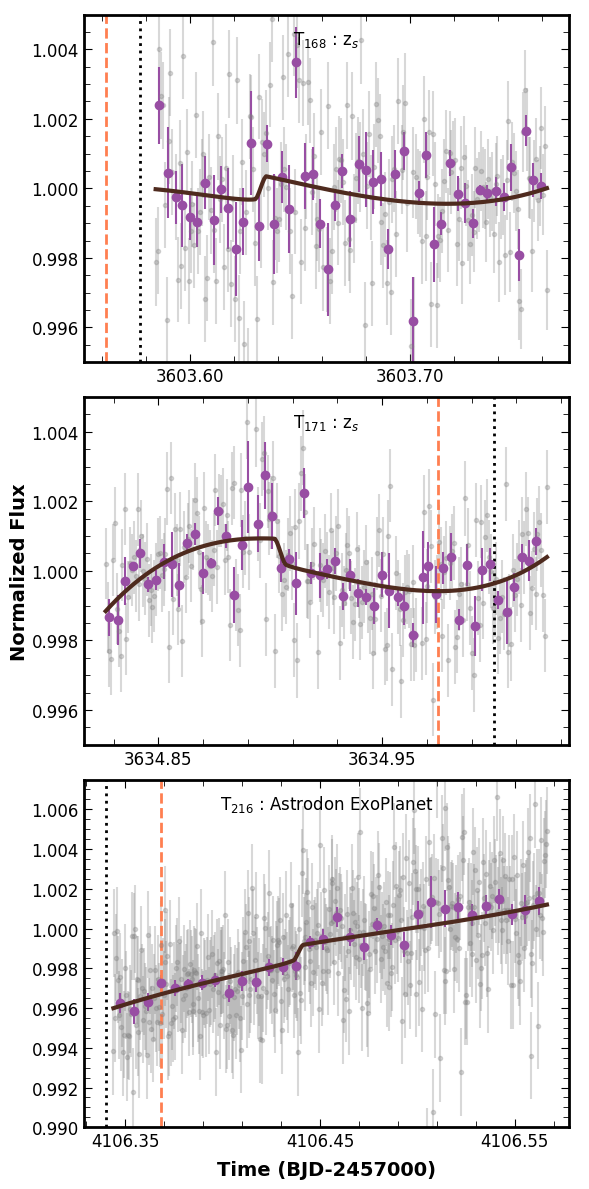}
    \caption{Partial transit detections with \textit{LCOGT}, labeled with the transit number and observation filter. The raw data is shown in gray and 10-minute bins are shown in purple for clarity. The best fit transit is shown as the brown solid line, the predicted transit midpoint (assuming a linear ephemeris) is shown as the black dotted line and the measured transit midpoint is shown as the pink dashed line.}
    \label{fig:partials}
\end{figure}

We fit the parameters in an MCMC framework using \texttt{emcee} \citep{emcee} with 100 walkers for 100,000 steps and a 20\% burn-in. We show the resulting fits in Figure \ref{fig:partials} and list the fit parameters in Table \ref{tab:partials}.

\begin{table*}[]
    \centering
    \caption{\planetname\ partial transit parameters}
    \begin{tabular}{lccc}
    \hline
    \hline
    Description & Parameter & Prior$^\alpha$ & Value\\
    \hline
    \multicolumn{4}{c}{Transit timings}\\
    \hline
    Transit 168 & $T_{168}$ & $N(3603.58, 0.05)$ & $3603.5620^{+0.0039}_{-0.0029}$ \\
    Transit 171 & $T_{171}$ & $N(3635.00, 0.05)$ & $3634.9786^{+0.011}_{-0.0066}$ \\
    Transit 216 & $T_{216}$ & $N(4106.34, 0.05)$ & $4106.3637^{+0.0077}_{-0.014}$  \\
    \hline
    \multicolumn{4}{c}{Photometric Trend Line}\\
    \hline
    $T_{168}$ $z_s$ offset & $\gamma_{168}$ & ... & $0.00090^{+0.00034}_{-0.00034}$ \\
    $T_{168}$ $z_s$ linear & $\dot\gamma_{168}$ & ... & $-0.0074^{+0.01636}_{-0.01672}$ \\
    $T_{168}$ $z_s$ quadratic & $\ddot\gamma_{168}$ & ... & $-0.09^{+0.22}_{-0.21}$ \\
    $T_{168}$ $z_s$ cubic & $\dddot\gamma_{168}$ & ... & $0.54\pm0.79$ \\
    $T_{171}$ $z_s$ offset & $\gamma_{171}$ & ... & $-0.0011\pm0.00030$ \\
    $T_{171}$ $z_s$ linear & $\dot\gamma_{171}$ & ... & $0.073\pm0.013$ \\
    $T_{171}$ $z_s$ quadratic & $\ddot\gamma_{171}$ & ... & $-0.81\pm0.15$ \\
    $T_{171}$ $z_s$ cubic & $\dddot\gamma_{171}$ & ... & $2.55\pm0.48$ \\
    $T_{216}$ Astrodon Exoplanet offset & $\gamma_{AE}$ & ... & $-0.0028^{+0.00043}_{-0.00044}$ \\
    $T_{216}$ Astrodon Exoplanet linear & $\dot\gamma_{AE}$ & ... & $0.017^{+0.019}_{-0.018}$  \\
    $T_{216}$ Astrodon Exoplanet quadratic & $\ddot\gamma_{AE}$ & ... & $0.02^{+0.18}_{-0.19}$ \\
    $T_{216}$ Astrodon Exoplanet cubic & $\dddot\gamma_{AE}$ & ... & $-0.06^{+0.54}_{-0.53}$ \\
    \hline
    \multicolumn{4}{c}{Limb-darkening Coefficients}\\
    \hline
    linear $z_s$ & $g_{1,z_s}$ & $N(0.347, 0.1)$ & $0.335^{+0.097}_{-0.098}$ \\
    quadratic $z_s$ & $g_{2,z_s}$ & $N(0.177, 0.05)$ &  $0.175^{+0.050}_{-0.049}$ \\
    linear Astrodon Exoplanet & $g_{1,AE}$ & $N(0.376, 0.1)$ & $0.38\pm0.10$\\
    quadratic Astrodon Exoplanet & $g_{2,AE}$ & $N(0.175, 0.05)$ & $0.18\pm0.05$ \\
    \hline
    \multicolumn{4}{l}{$\alpha$ $N(a,b)$ indicates a normal distribution centered at $a$ with a standard deviation of $b$.}
    \end{tabular}
    \label{tab:partials}
\end{table*}

All transit times can be seen in Figure~\ref{fig:ttv}. The TTV is not visible from the \tess{} data alone, but clear when including all ground-based data. Because of strong stellar variability, it is possible we are underestimating uncertainties on some transit times, especially for the partials. However, the TTV is visible even if we exclude the partial transits. Further, the semi-amplitude is $\gtrsim30$\,minutes, which is challenging to explain by data quality, spots, or imperfectly corrected variability alone \citep{LopezMurillo2026}. Due to the sparse sampling and only using the full transit observations in the \texttt{Juliet} analysis, the calculated planet period may be off. Further transit observations are needed to constrain the true planet period and TTV amplitude.

\subsection{Rossiter-McLaughlin Fit}

Following \cite{soysauce1}, we simultaneously fit the \maroonx\ velocities using an updated version of the RM implementation in \texttt{MISTTBORN} \citep[see][]{MISTTBORN, Johnson2022}, and the LCO photometric data using a \texttt{BATMAN} and polynomial variability model. The RM uses the methodology in \cite{Hirano_2011} and \cite{Addison_2013} to produce a model of the RM using the analytical functions of the rotational broadening (\vsini) and intrinsic width of the Gaussian line profile of individual surface elements ($v_{int}$) and taking into account the change in flux from the transit. This allowed us to fit the red and blue \maroonx\ data and photometric data with common transit parameters but unique trend lines and limb-darkening parameters.

We fit for 25 parameters in total. For the planet parameters, we fit the dataset with a common transit midpoint ($T_0$), period ($P$), impact parameter ($b$), stellar density ($\rho_*$), and planet-to-star radius ratio ($R_P/R_*$). For the RV data, we fit each channel with a common sky-projected obliquity angle ($\lambda$), \vsini, and $v_{int}$. For each dataset, we fit for unique limb-darkening parameters ($q_1$, $q_2$) and trend lines. For the RVs, we fit each channel with an RV slope ($\dot\gamma$) and an RV offset ($\gamma$), and for the photometry, we fit for a third-order polynomial with the first data point redefined to be t=0. We observed a flare-like event in the photometry $\sim2.5$hours before midpoint, motivating us to include a flare model following \cite{Davenport2014} described by the flare midpoint ($t_{flare}$), amplitude ($A$), and the full time width at half the maximum flux ($t_{1/2}$).  

We initialized the transit parameters and applied Gaussian priors based on the photometric data fit (Section \ref{sec:juliet}). As above, we placed Gaussian priors on the limb-darkening parameters derived from \texttt{LDTK}. We also placed a Gaussian prior on $v_{int}$ drawn from the instrumental width and an estimate of the macroturbulence from \cite{Brewer_2016}.

\begin{figure*}
    \centering
    \includegraphics[width=0.99\linewidth]{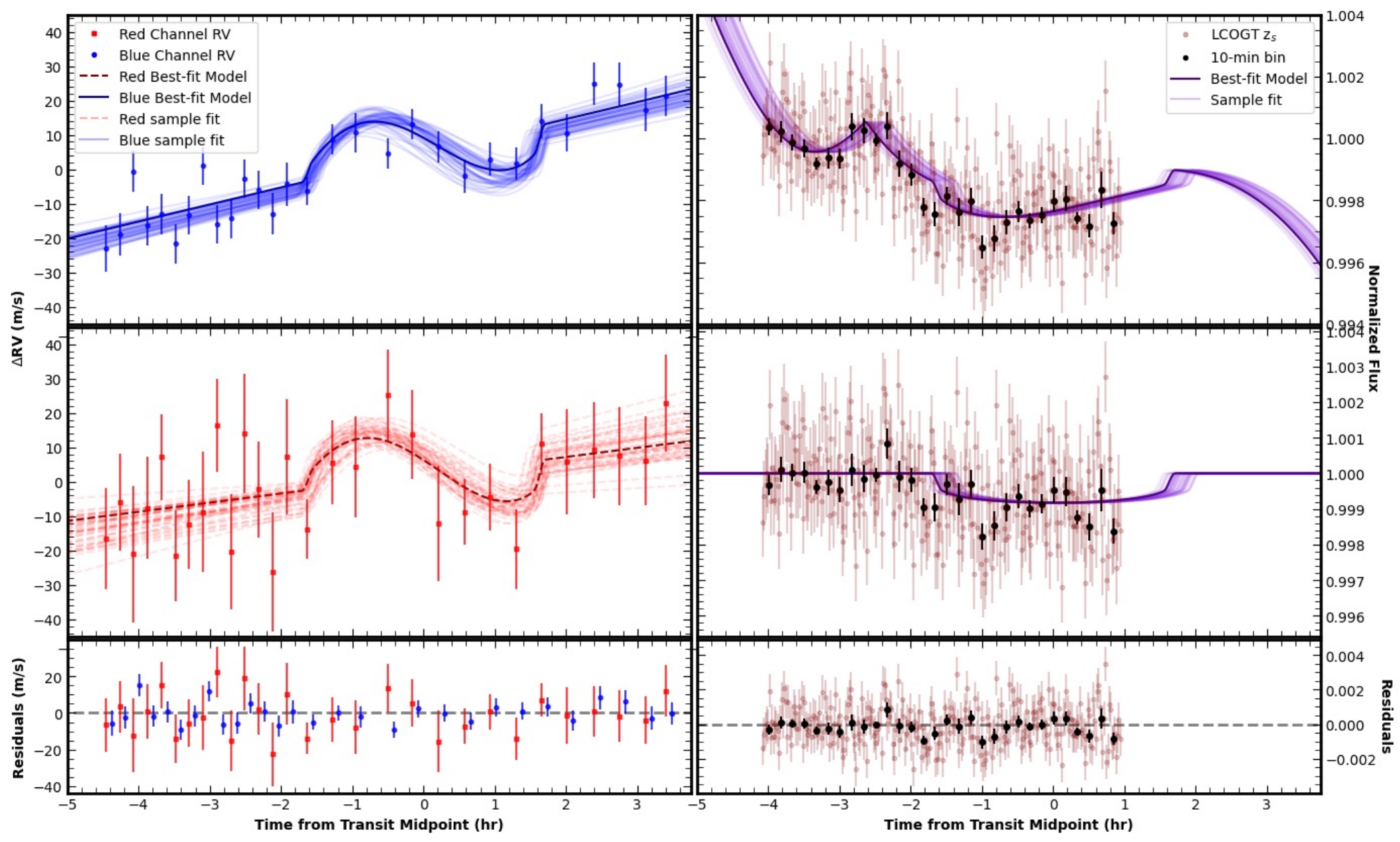}
    \caption{Left) \maroonx\ relative radial velocities from the blue channel (top, blue circles) and red channel (middle, red squares). The best-fit RM model and stellar RV trend lines are shown as the opaque solid blue and dashed red lines, with 50-sample fits pulled from the posterior shown as the translucent blue and red lines. The bottom plot shows the residuals for each channel, with a 5-minute shift applied to the blue channel for clarity. Right) \textit{LCOGT} photometric data (brown) with 10-min bins (black) shown for clarity. The best-fit model is shown as the opaque purple line, with 50-sample fits pulled from the posterior shown as the translucent purple lines. The top panel shows the raw data, the middle panel shows the data and model after the stellar signal and flare have been removed, and the bottom plot shows the residuals.}
    \label{fig:rm}
\end{figure*}

We used \texttt{emcee} to fit the parameters in an MCMC framework with 100 walkers, 150,000 steps, and a 20\% burn-in. The total run time was more than 50 times the autocorrelation time, sufficient for convergence. We list the priors and best-fit parameters in Table \ref{tab:rm} and show the best-fit model in Figure \ref{fig:rm}.

\begin{table*}[]
    \centering
    \caption{Rossiter-McLaughlin fit parameters and priors}
    \begin{tabular}{lccc}
    \hline
    \hline
    Description & Parameter & Prior$^\alpha$ & Value\\
    \hline
    \multicolumn{4}{c}{Transit Parameters}\\
     \hline
    transit midpoint & $T_0$ (BJD-2457000) & $N(4033.021, 0.1)$ & $4032.9920^{+0.0050}_{-0.0065}$\\
    planet-to-star radius ratio & $R_P/R_*$ & $N(0.031, 0.001)$ & $0.0289\pm0.0011$ \\
    impact parameter & $|b|$ & $N(0.53, 0.15)$ & $0.58\pm0.10$ \\
    stellar density & $\rho_*$ ($\rho_\odot$) & $N(1.07, 0.34)$, $\rho_*>0$ & $0.99\pm0.20$ \\
    orbital period & $P$ (days) & $N(10.474230, 0.00004)$ & $10.4742302\pm0.000044$ \\
    \hline
    \multicolumn{4}{c}{Rossiter-McLaughlin Parameters}\\
     \hline
     sky-projected obliquity & $|\lambda|$ ($\degree$) & ... & $18.21^{+11.48}_{-12.35}$ \\
     rotational broadening & \vsini (\kms) & $N(19, 1)$ & $18.09\pm0.97$ \\
     intrinsic width & $v_{int}$ (\kms) & $N(5.0, 0.5)$ & $4.99\pm0.50$ \\
    \hline
    \multicolumn{4}{c}{RV Trend Line Parameters}\\
     \hline
     \maroonx$_{blue}$ offset & $\gamma_{blue}$ & ... & $-0.0194^{+0.0019}_{-0.0020}$ \\
     \maroonx$_{blue}$ slope & $\dot\gamma_{blue}$ & ... & $0.126\pm0.012$ \\
     \maroonx$_{red}$ offset & $\gamma_{red}$ & ... & $-0.0125\pm0.0046$ \\
     \maroonx$_{red}$ slope & $\dot\gamma_{red}$ & ... & $0.0695\pm0.025$ \\
     \hline
    \multicolumn{4}{c}{Photometric Trend Line Parameters}\\
    \hline
     $z_s$ offset & $\gamma_{z_s}$ & ... & $0.00070^{+0.00038}_{-0.00051}$ \\
     $z_s$ linear & $\dot\gamma_{z_s}$ & ... & $-0.0904^{+0.015}_{-0.014}$ \\
     $z_s$ quadratic & $\ddot\gamma_{z_s}$ & ... & $0.64\pm0.15$ \\
     $z_s$ cubic & $\dddot\gamma_{z_s}$ & ... & $-1.24^{+0.46}_{-0.45}$ \\
     \hline
     \multicolumn{4}{c}{Flare parameters}\\
     \hline
     midpoint & $t_{flare}$ (BJD-2457000) & ... & $4032.8861^{+0.0028}_{-0.0026}$ \\
     amplitude & $A$ & ... & $0.00307\pm0.00029$ \\
     full time width at half the maximum flux & $t_{1/2}$ (days) & ... & $0.084^{+0.024}_{-0.019}$ \\
     \hline
     \multicolumn{4}{c}{Limb-darkening Coefficients}\\
     \hline
     linear \maroonx$_{blue}$ & $g_{1,blue}$ & $N(0.475, 0.1)$ & $0.49\pm0.10$ \\
     quadratic \maroonx$_{blue}$ & $g_{2,blue}$ & $N(0.178, 0.05)$ & $0.178\pm0.049$  \\
     linear \maroonx$_{red}$ & $g_{1,red}$ & $N(0.363, 0.1)$ & $0.370^{+0.099}_{-0.098}$ \\
     quadratic \maroonx$_{red}$ & $g_{2,red}$ & $N(0.173, 0.05)$ & $0.174\pm0.049$ \\
     linear $z_s$ & $g_{1,z_s}$ & $N(0.318, 0.1)$ & $0.320\pm0.099$ \\
     quadratic $z_s$ & $g_{2,z_s}$ & $N(0.17, 0.05)$ & $0.171\pm0.050$ \\
     \hline
     \multicolumn{4}{l}{$\alpha$ N(a,b) indicates a normal distribution centered at a with a standard deviation of b.}
    \end{tabular}
    \label{tab:rm}
\end{table*}

Due to the scatter, the photometric light curve provides only weak additional constraints on the transit timing. Removing the LCO data and fitting just the RM signal, we find consistent fit parameters and $\lambda$ values.

We summarize the fit parameters in Table~\ref{tab:rm}. The planet is consistent with alignment with the host star ($|\lambda|=18 \pm 12\degree$). The \vsini\ of 19\kms also suggests an aligned system (equatorial velocity is $20.1\pm0.5$\kms). We combine these to compute the true three-dimensional obliquity, $\psi$. This is also consistent with alignment, yielding $\psi=23\pm11^\circ$ or a 95\% upper limit of $38^\circ$. Uncertainties are relatively large compared to other spin-orbit alignment measurements, but this is largely due to uncertainty about $b$, and hence should improve with additional high-precision transits. 

\subsection{Doppler Tomography}

Given the stellar variability of the target, we also investigated the possibility of performing a Doppler Tomographic (DT) analysis of the spectra. DT resolves the distortion in the stellar line profile caused by the planet’s shadow, and hence is less impacted by stellar signals than an RM analysis.

Unfortunately, the star's moderate rotational velocity (\vsini$=20$km s$^{-1}$) means the expected shadow is smaller than the line profile (0.5\kms\ compared to 3-4\kms), weakening the signal and providing poor velocity resolution of any motion by the planet. Combined with a shallow transit depth ($\sim$900 ppm), the SNR required for a clear detection is expected to be $>$500 (per exposure), while our spectra hit a peak SNR of $\simeq$150. A cursory examination of the stacked template (SNR$\simeq$300) with a least-squares deconvolution code \citep{Kochukhov2010} confirmed that the expected signal is far below detection. 

\subsection{Mass limit}
RV campaigns of young active stars to recover planet masses are complicated by young stellar activity \citep{Blunt2023}. Cross-instrument offsets can also be significant (10s or 100s of \mps) compared to the planetary signal. While intensive high-precision campaigns from a single instrument is necessary to achieve tight mass-limits on the youngest systems \citep[e.g.][]{Donati2025}, we can use the inhomogeneous and sparsely sampled spectra of the system to rule out stellar mass objects ($>$13\,$M_J$) where the expected signal is $>1000$\,\mps.

\begin{figure}
    \centering
    \includegraphics[width=0.99\linewidth]{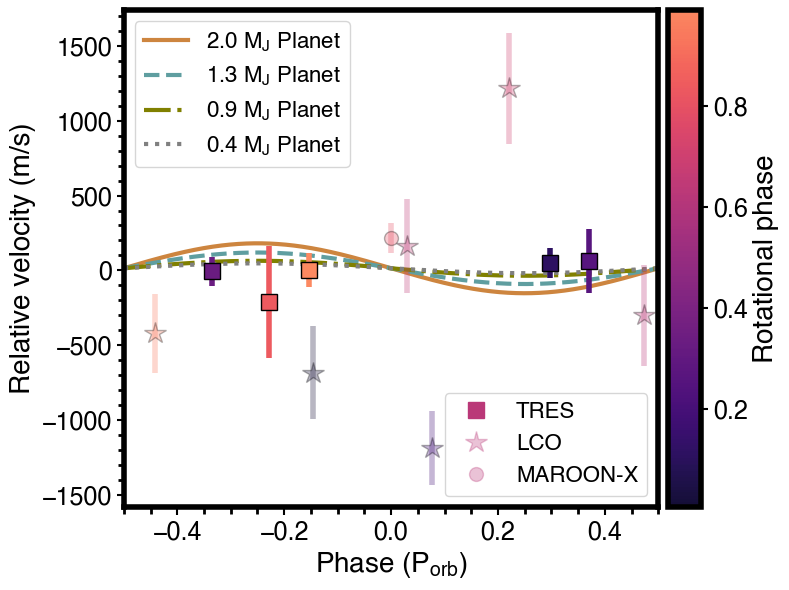}
    \caption{Relative radial velocities of \starname\ colored by the observation's phase in the stellar rotation (arbitrary zero-point) and phased to \planetname's orbital period. The median fit (dotted line) of the TRES (squares) observations is shown, as well as the 1, 2, and 3$\sigma$ mass limits and the LCO (stars) and MAROON-X (circle) epochs for reference.}
    \label{fig:rv}
\end{figure}

Following \cite{Barber2024_iras}, we fit for the semi-amplitude of the RVs ($K$) and a velocity offset ($\gamma$). We initially included an extra jitter term ($\sigma_J$) but found $\sigma_J$ was unbound, so we opted to remove it from the final fit. We normalized the \textit{TRES}, \textit{LCO}, and \textit{MAROON-X} epochs into relative radial velocities by subtracting the median value of the 12 epochs. For \textit{MAROON-X}, we adopt the median absolute RV across both channels. We find an upper mass limit of 1.1 $M_J$ at 95\%. Isolating the \textit{TRES} epochs, we find an upper limit of 1.3 $M_J$ at 95\%, well below a stellar-mass object. We show the resulting fit in Figure \ref{fig:rv} and list the RVs in Table \ref{tab:rv}. Instrumental offsets might explain the variation in mass, but even considering TRES-only RVs still provide limits in the planetary regime.

\begin{table}[]
    \centering
    \begin{tabular}{lcccc}
    \hline
    Epoch & Absolute & Relative & $\sigma$ & Facility \\
    (BJD) & (\kms) & (\kms) & (\kms) & \\
    \hline
    2460407.69 & 12.262 & 0.049 & 0.098 & TRES\\
    2460412.66 & 12.000 & -0.213 & 0.375 & TRES\\
    2460722.67 & 12.276 & 0.063 & 0.217 & TRES\\
    2460725.77 & 12.209 & -0.005 & 0.097 & TRES\\
    2460727.67 & 12.218 & 0.005 & 0.114 & TRES\\
    2461002.40 & 11.028 & -1.185 & 0.247 & LCO \\
    2461003.92 & 13.431 & 1.217 & 0.372 & LCO \\ 
    2461006.57 & 11.913 & -0.300 & 0.338 & LCO\\
    2461007.45 & 11.792 & -0.422 & 0.262 & LCO\\
    2461010.55 & 11.529 & -0.684 & 0.311 & LCO\\
    2461012.38 & 12.378 & 0.165 & 0.315 & LCO\\
    2461033.03 & 12.43 & 0.216 & 0.1 & MAROON-X\\
    \hline
    \end{tabular}
    \caption{Absolute and relative radial velocities of \starname.}
    \label{tab:rv}
\end{table}

\subsection{Candidate \starname\ c}\label{sec:planetc}
We detected an additional candidate signal in the \starname\ system at 7.9 days which passed our SNR requirements (Section~\ref{sec:transit_search}). Upon further inspection of the light curve, we deduced this was a likely half period alias, as there were no convincing transit signals at or near the expected timings in Sector 47 which matched the depths of the signals in Sector 20.

We fit for the planet parameters using a \texttt{BATMAN} transit model and a linear trend line as above. Due to the stellar variability, we cut out a 12-hour window centered at each of the two expected transit timings based on the detection ephemeris ($T_0=1848.07$ and $T_1=1863.94$). We fit for the time of inferior conjunction ($T_c$), period ($P$), the planet-to-star radius ratio ($R_P/R_*$), impact parameter ($b$), and stellar density ($\rho_*$). For the stellar variability, we fit for a slope ($\dot\gamma$) and y-offset ($\gamma$). Each transit was fit with a unique trend line but common planet parameters.

The parameters were allowed to float with physical priors, with the exception of $\rho_*$ and the limb-darkening parameters, which we set priors on based on the \texttt{Juliet} fit of \planetname\ (Section \ref{sec:juliet}).

Using an MCMC framework with \texttt{emcee}, we fit for the 11 parameters with 100 walkers and 100,000 steps. We show the resulting fit in the top-right of Figure \ref{fig:phased} and list the fit parameters in Table \ref{tab:cparams}. 

With only two transits, the properties of planet c are poorly constrained. As we discuss in our false positive analysis section below, the period ratio is important for interpreting the TTV signal seen in the b planet. 

\begin{table*}[]
    \centering
    \caption{Transit fit parameters of \starname\,c}
    \begin{tabular}{lccc}
    \hline
    \hline
    Description & Parameter & Prior$^\alpha$ & Value\\
    \hline
    \multicolumn{4}{c}{Transit Parameters}\\
     \hline
    planet-to-star radius ratio & $R_P/R_*$ & ... & $0.026^{+0.0022}_{-0.0020}$ \\
    time of inferior conjunction & $T_c$ (BJD - 2457000) & ... & $1848.041^{+0.033}_{-0.015}$ \\
    impact parameter & $|b|$ & $U(0,1)$ & $0.38^{+0.23}_{-0.25}$ \\
    stellar density & $\rho_*$ ($\rho_\odot$) & $N(1.07, 0.32)$, $\rho_*>0$ & $1.11^{+0.26}_{-0.27}$ \\
    orbital period & $P$ & ...  & $15.905^{+0.016}_{-0.044}$ \\
    \hline
    \multicolumn{4}{c}{Variability Parameters}\\
     \hline
    flux offset $T_0$ & $\gamma_0$ & ... & $-0.00108\pm0.00013$ \\
    flux slope $T_0$ & $\dot\gamma_0$ & ... & $-0.00945\pm0.00038$ \\
    flux offset $T_1$ & $\gamma_1$ & ... & $0.00551^{+0.00011}_{-0.00012}$ \\
    flux slope $T_1$ & $\dot\gamma_1$ & ... & $-0.00760\pm0.00034$ \\
     \hline
    \multicolumn{4}{c}{Limb-darkening Coefficients}\\
     \hline
     linear \tess\ & $g_{1,\tess}$ & $N(0.376, 0.1)$ & $0.375^{+0.100}_{-0.098}$ \\
     quadratic \tess\ & $g_{2,\tess}$ & $N(0.175, 0.05)$ & $0.175^{+0.050}_{-0.049}$ \\
     \hline
     \hline
    \multicolumn{4}{c}{Derived Parameters}\\
    \hline
    planet radius & $R_P$ ($R_\oplus$) & ... & $3.03^{+0.30}_{-0.26}$\\
    transit duration (first-to-fourth contact) & $t_{dur}$ (days) & ... & $0.17^{+0.012}_{-0.016} $\\
    semi-major axis to stellar radius ratio & $a/R_*$ & ... & $27.57_{-2.47}^{+1.98} $\\
    inclination & $i(\degree)$ & ... & $90.01^{+1.05}_{-1.06}$\\
    equilibrium temperature$^{\dagger}$ & $T_{eq}$ (K)& ... & $816.71_{-28.70}^{+39.73}$\\
    \hline
    \multicolumn{4}{l}{$\alpha$ $U(a,b)$ indicates a uniform distribution from $a$ to $b$.}\\
    \multicolumn{4}{l}{\,\,\,\,\,$N(a,b)$ indicates a normal distribution centered at $a$ with a standard deviation of $b$.}\\
    \multicolumn{4}{l}{$\dagger$ assuming zero albedo}
    \end{tabular}
    \label{tab:cparams}
\end{table*}

\subsection{Transit timing analysis}

As we show in Figure~\ref{fig:ttv}, planet b exhibits transit timing variations of $\simeq$30m from the linear ephemeris. This is too large to be explained by the star's stellar activity \citep[e.g.][]{LopezMurillo2026}, although activity combined with a large number of partial transits may make the amplitude look larger than it is. The TTV period and amplitude are sensitive to the orbital period of planet c due to the system's proximity to 3:2 mean motion resonance. At the adopted period (15.905\,d, $\simeq$1.2\% outside exact resonance, see Section \ref{sec:planetc}), the predicted super-period is 430 days, with a 1$\sigma$ range of 397--555\,days. A simple sine-curve fit to the data favors a period of 380\,days, within 1.5$\sigma$ of the expected range. 

We explored this further using the code \texttt{TTVFast} \citep{Deck2014}. Because of the relatively small number of data points, we restricted the analysis to circular orbits and masses $5-40\,M_\oplus$ for both planets. We had the period and $T_0$ of both planets evolve under Gaussian priors from the analysis above, although the existence of the TTV means that these uncertainties are likely underestimated. We performed a Maximum A Posteriori (MAP) optimization using \texttt{scipy}'s differential evolution algorithm, minimizing the sum of a chi-squared likelihood and Gaussian prior terms. Consistent with the basic calculation above, the \texttt{TTVFast} fit prefers a period for planet c on the high end of the fit (15.93\,days or 2$\sigma$ from the nominal fit), to reproduce the periodicity in the data, and a high mass for planet c ($\simeq30M_\oplus$) and low for planet b (5-10$M_\oplus$) to compensate for moving the period further from resonance. The best-fit has a reduced $\chi^2$ of 2. Eccentricity and a wider range of parameters would likely improve this fit, but require more transits of both planets to narrow down the range of solutions.

\section{False Positive Analysis} \label{sec:fp}

\subsection{Planet b}
Statistical false-positive assessments \citep[such as \texttt{TRICERATOPS};][]{triceratops_code,Giacalone2021} can be complicated by the presence of young stellar noise and the need to detrend the light curve. These frameworks are sensitive to the method and quality of the detrending and any residual noise in the light curve. However, they can be useful for ruling out certain kinds of false positives that impact both young and old systems, provided the light curve is of sufficient quality. Using \texttt{TRICERATOPS} with the GP-detrended \texttt{Juliet} light curve and speckle imaging of the star, we find a false positive probability (FPP) of $<0.009$ for planet b. While this is sufficient to validate (FPP $<$0.015), we can also use the follow-up observations of the system as an additional check.

The transit is recovered on-source from the ground (with apertures as small at 4" for photometry). Speckle imaging rules out any companion with $\Delta$m$<$4.5 at 0.1" separation. Any unresolved companion of this magnitude and proximity would manifest as a second set of lines in our stacked (SNR$>300$) MAROON-X spectra or any of the other high-resolution spectra. Most importantly, the \maroonx\ RM confirms the transiting object must be orbiting \starname, otherwise the spectral line shifts in transit would not be visible. Further, those shifts would show strong chromaticity, as any unseen star would not have identical \teff{} and have $\Delta RV=0$ over multiple years of epochs. Comparing the maximum log-likelihoods of the RM model and a trend-line-only (no RM signal) model, we find a $\Delta\chi^2\simeq23$, corresponding to an SNR\,$\simeq5$, favoring the RM model and a transit signal. The presence of a TTV also provides independent validation, as the only way to create such large timing variations on a stellar companion is if there's a stellar-mass perturber, both of which would be easily detected in our suite of imaging and spectra.

\subsection{Planet c}

While the ephemeris of the second planet signal is not precise enough to confirm in current epochs, the TTV detected in \planetname\ suggests planet c is likely real. The transit timing variations seen in the transits of \planetname\ strongly suggests a second co-planar object is in the system. TTVs are commonly seen in young systems, especially those in or near mean motion resonance \citep{Dai2024,LopezMurillo2026}. The candidate signal is just outside mean motion resonance with \planetname\ (1.51:1 or approximately 3:2) and in a period ratio commonly observed in \kepler\ multi-planet systems \citep{Fabrycky2014}. The likelihood the second signal lands in a first order resonance with \planetname\ by chance is small.

Using \texttt{TRICERATOPS}, we find a FPP of $<$0.001, sufficient for statistical validation. While this is further evidence for \starname\,c being a real planet, we are not comfortable claiming validation with just two transit signals. Unfortunately, the data gap in the second sector observations of \starname\ coincides with the expected transit timing of the candidate. Due to the long period, imprecise ephemeris, and no scheduled \tess\ follow-up, we cannot definitively confirm the signal and \starname\ c remains a planet candidate pending additional transits.

\section{Summary and Conclusions}\label{sec:discussion}

We have characterized and validated the planet \planetname{} as well as its host association, Alessi 84. Our major findings are:
\begin{itemize}
    \item Alessi 84 is 135$\pm$10\,Myr based on the CMD, rotation, and variability of likely members. 
    \item \starname's rotation and lithium yield a consistent young age, which when combined with the star's kinematics, confirms \starname\ is part of Alessi 84.
    \item RM observations of \starname{} indicate \planetname{} is aligned with its host star ($|\lambda|<36^\circ$ and $\psi<38^\circ$ at 95\%). Most of the uncertainty is due to uncertainty in the planet's impact parameter. 
    \item \planetname{} exhibits $\gtrsim 30$\,minute transit timing variations. Combined with the RM data and the wealth of other follow-up, this validates the signal as planetary.
    \item We detect an additional planet candidate nearby a 3:2 period ratio. The presence of the significant TTV in the b planet suggests this planet is real. However, with only two transits we consider it a candidate pending additional observations.
\end{itemize}

TIC 150070085 joins the sparsely sampled population of near 100 Myr transiting planet systems. Closely aged systems Kepler-1928 and Kepler-970 \citep[105 Myr;][]{Barber2022_kepler} and K2-284 \citep[120 Myr;][]{David2018_k2284} are much smaller planets discovered by \kepler\ around much fainter stars. The \tess\ discovered systems with precise age determinations closest to TIC 150070085 are TOI-1860 \citep[$133\pm26$ Myr;][]{Giacalone2022} and TOI-1224 \citep[$210\pm27$ Myr;][]{Thao2024_1224}, providing an important data point in mapping the evolution of planets over time. 

\begin{figure*}
    \centering
    \includegraphics[width=0.99\linewidth]{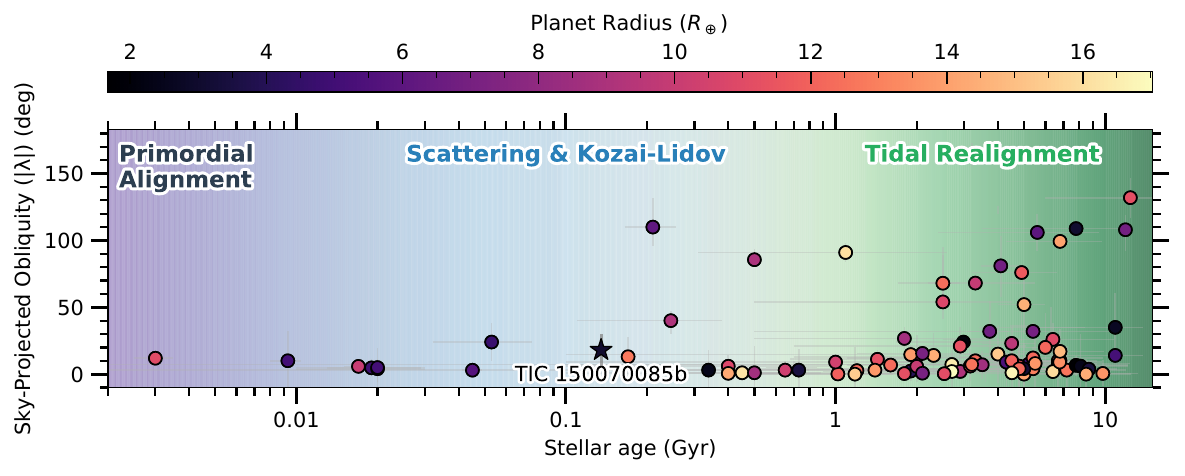}
    \caption{Distribution of sky-projected spin-orbit alignment estimates ($\lambda$) for stars with reported ages. Targets are color-coded by planetary radius. \starname{} is shown as a star. \starname{}'s general alignment is consistent with younger $<100$\,Myr systems, but more targets in the 0.1-1\,Gyr range are needed to determine when misalignments start to occur.  }
    \label{fig:soysauce}
\end{figure*}

\starname{} is also one of the modest number of young planets with an obliquity measurement from Rossiter-McLaughlin or Doppler tomography. \starname{} follows the pattern of other young planets that are aligned or with small misalignments ($\lesssim 20^\circ$) (Figure~\ref{fig:soysauce}). While most old systems are also aligned, there's a clear population of misalignments not similarly seen in systems $\lesssim$200\,Myr. More observations are needed for this to be statistically significant, especially since some young systems show evidence of misalignment \citep{Alves2025}.

A strong TTV ($\simeq$30\,m) offers an excellent opportunity to characterize the mass and eccentricity of the system. However, this requires additional high-precision observations of \starname\ to confirm the TTV amplitude and recover planet c. The ephemeris for candidate planet c has decayed to a level where this is impractical. The best opportunity to confirm the signal would be for \tess\ to reobserve \starname\ in a future mission extension. With no current such plans, the only other option may be a multi-day stare with \textit{CHEOPS}. 

\section*{Acknowledgments}
The authors wish to thank Halee and Bandit for their continued commitment to supporting scientific progress and the scientists behind it. MGB was supported by the NSF Graduate Research Fellowship (DGE-2040435). AWM was supported by the NSF CAREER program (AST-2143763) and a grant from NASA's Exoplanet Research Program (XRP 80NSSC21K0393). LJB was supported by the Chancellor's Science Scholars Program\footnote{\url{https://chancellorssciencescholars.unc.edu/}}. AWB was supported by the NSF Graduate Research Fellowship (DGE-2439854). AWB thanks the LSST-DA Data Science Fellowship Program, which is funded by LSST-DA, the Brinson Foundation, the WoodNext Foundation, and the Research Corporation for Science Advancement Foundation; his participation in the program has benefited this work. 

We acknowledge financial support from the Agencia Estatal de Investigaci\'on of the Ministerio de Ciencia e Innovaci\'on MCIN/AEI/10.13039/501100011033 and the ERDF “A way of making Europe” through projects PID2021-125627OB-C32 and PID2024-158486OB-C32.

Funding for the \tess\ mission is provided by NASA’s Science Mission Directorate. We acknowledge the use of public \tess\ data from pipelines at the \tess\ Science Office and at the \tess\ Science Processing Operations Center. Resources supporting this work were provided by the NASA High-End Computing (HEC) Program through the NASA Advanced Supercomputing (NAS) Division at Ames Research Center for the production of the SPOC data products. \tess\ data presented in this paper were obtained from the Mikulski Archive for Space Telescopes (MAST) at the Space Telescope Science Institute.

This research has made use of the Exoplanet Followup Observation Program \citep[ExoFOP;][]{EXOFOP} website, which is operated by the California Institute of Technology, under contract with the National Aeronautics and Space Administration under the Exoplanet Exploration Program.

This work has made use of data from the European Space Agency (ESA) mission Gaia (\url{https://www.cosmos.esa.int/gaia}), processed by the Gaia Data Processing and Analysis Consortium (DPAC; \url{https://www.cosmos.esa.int/web/gaia/dpac/consortium}). Funding for the DPAC has been provided by national institutions, in particular the institutions participating in the Gaia Multilateral Agreement.

This work makes use of observations from the LCOGT network. 

\startlongtable
\begin{longrotatetable}
    \begin{deluxetable*}{ccccccccccccccc}\label{tab:group}
    \tablehead{
    \colhead{Gaia DR3} & \colhead{$\alpha$} & \colhead{$\delta$} & \colhead{Gmag} & \colhead{$B_p$-$R_p$} &  \colhead{$RV$\tablenotemark{a}} & \colhead{$\sigma_{RV}$\tablenotemark{a}} & \colhead{$P_{rot}$\tablenotemark{b}} & \colhead{$\sigma_{Prot}$\tablenotemark{b}} & \colhead{FF\tablenotemark{c}} & \colhead{Alessi\tablenotemark{d}} & \colhead{Theia\tablenotemark{e}} & \colhead{CWNU\tablenotemark{f}} & \colhead{OCSN\tablenotemark{g}} & \colhead{UPK\tablenotemark{h}}\\
    \colhead{} & \colhead{(deg, J2016)} & \colhead{(deg, J2016)} & \colhead{(mag)} & \colhead{(mag)}  & \colhead{(km/s)} & \colhead{(km/s)} & \colhead{(days)} & \colhead{(days)} & \colhead{}  & \colhead{84} & \colhead{214} & \colhead{1128} & \colhead{304} & \colhead{343} 
    }
    \startdata
    986757333119241216 &  110.1432 &   53.547 & 10.99 &  0.78 &  12.43 &          0.74 &      2.89 &            0.04 &  Y &         Y &         Y &         Y &        Y &       Y \\
 986520835040628608 &  110.0536 &  52.8409 & 16.05 &  2.51 &    ... &           ... &       ... &             ... &  Y &         Y &         Y &         Y &        Y &       Y \\
 989812979012735744 &  109.7498 &  56.8705 & 17.34 &   2.9 &    ... &           ... &       ... &             ... &  Y &         Y &         Y &         Y &        Y &       Y \\
 988631759926303104 &  111.3347 &  55.1176 &  9.59 &  0.56 &   9.31 &          1.08 &       ... &             ... &  Y &         Y &         Y &         Y &        Y &       Y \\
 988885575313972352 &  111.0834 &  56.0102 & 16.37 &  2.66 &    ... &           ... &       ... &             ... &  Y &         Y &         Y &         Y &        Y &       Y \\
 988644954065633024 &  111.2642 &  55.3967 & 16.76 &   2.7 &    ... &           ... &     0.781 &           0.005 &  Y &         Y &         Y &         Y &        Y &       Y \\
 986919098767707520 &  111.3262 &  53.8105 & 10.03 &  0.61 &  13.66 &          1.01 &      1.04 &            0.01 &  Y &         Y &         Y &         Y &        Y &       Y \\
 987407002755987328 &  109.6884 &  55.6075 & 11.98 &  0.98 &  10.22 &          0.64 &       ... &             ... &  Y &         Y &         Y &         Y &        Y &       Y \\
 989060019706031872 &  110.4508 &  56.7477 & 12.71 &  1.17 &   7.03 &          1.81 &      8.45 &            0.29 &  Y &         Y &         Y &         Y &        Y &       Y \\
 987389483585544192 &  110.1364 &  55.4082 & 15.33 &  2.16 &  15.12 &         11.86 &       ... &             ... &  Y &         Y &         Y &         Y &        Y &       Y \\
 987204078437237376 &  109.4654 &  54.0602 & 17.36 &  3.02 &    ... &           ... &       ... &             ... &  Y &         Y &         Y &         Y &        Y &       Y \\
 988902858262422912 &   110.336 &  55.8513 & 17.49 &   2.9 &    ... &           ... &       ... &             ... &  Y &         Y &         Y &         Y &        Y &       Y \\
 987335607515837568 &   110.607 &  55.1635 & 10.92 &  0.77 &  11.17 &          0.65 &       3.0 &            0.04 &  Y &         Y &         Y &         Y &        Y &       Y \\
 988611899997524736 &  111.6568 &  54.8218 & 12.98 &  1.26 &   9.46 &          1.13 &      8.16 &            0.27 &  Y &         Y &         Y &         Y &        Y &       Y \\
 987034959805184640 &  112.0379 &  54.2194 &  9.92 &  0.59 &  10.76 &          0.66 &       ... &             ... &  Y &         Y &         Y &         Y &        Y &       Y \\
 987331518706989568 &  110.2701 &  55.1065 &  9.71 &  0.55 &  11.35 &          0.57 &       ... &             ... &  Y &         Y &         Y &         Y &        Y &       Y \\
 988842041525566720 &  110.7498 &   55.481 &  8.86 &  0.26 &   6.83 &          0.77 &       ... &             ... &  Y &         Y &         Y &         Y &        Y &       Y \\
 987337939681876352 &  110.6775 &  55.2906 & 12.53 &  1.11 &  11.27 &           1.1 &       ... &             ... &  Y &         Y &         Y &         Y &        Y &       Y \\
 988902853966489856 &  110.3365 &  55.8507 & 17.49 &  2.97 &    ... &           ... &       ... &             ... &  Y &         Y &         Y &         Y &        Y &       Y \\
 988841354330807680 &  110.5762 &  55.4705 & 15.22 &  2.15 &   14.8 &         11.71 &      5.97 &            0.15 &  Y &         Y &         Y &         Y &        Y &       Y \\
 987088629716470912 &  110.7645 &  54.4963 & 14.74 &  1.95 &    5.4 &          5.76 &     11.73 &            0.79 &  Y &         Y &         Y &         Y &        Y &       Y \\
 988827610435458816 &  110.7244 &  55.3408 &  7.56 &  0.01 &  11.76 &          0.74 &       ... &             ... &  Y &         Y &         Y &         Y &        Y &       Y \\
 986985550500089728 &  110.9233 &  54.0915 &  8.38 &  0.08 &  18.03 &          3.03 &       ... &             ... &  Y &         Y &         Y &         Y &        Y &       Y \\
 988572832974864512 &  111.9627 &  54.9598 & 16.46 &  2.66 &    ... &           ... &     0.645 &           0.003 &  Y &         Y &         Y &         Y &        Y &       Y \\
 986999844152694016 &  111.8465 &  53.8487 & 16.91 &  2.86 &    ... &           ... &       ... &             ... &  Y &         Y &         Y &         Y &        Y &       Y \\
 987326158587778432 &  110.7169 &  55.2813 &  5.78 & -0.11 &    ... &           ... &       ... &             ... &  Y &         Y &         Y &         Y &        Y &       Y \\
 987412195372614656 &  109.7134 &  55.7299 & 16.03 &  2.49 &    ... &           ... &       ... &             ... &  Y &         Y &         Y &         Y &        Y &       Y \\
 987283930469428736 &  110.6553 &  54.6484 & 10.11 &  0.74 &  11.28 &          0.66 &       ... &             ... &  Y &         Y &         Y &         Y &        Y &       Y \\
 987092237488984448 &  111.0636 &  54.5402 &  8.90 &   0.3 &   9.84 &           0.8 &       ... &             ... &  Y &         Y &         Y &         Y &        Y &       Y \\
 987146010479417216 &  110.1014 &  53.8779 &  9.90 &  0.59 &  11.02 &          0.39 &       ... &             ... &  Y &         Y &         Y &         Y &        Y &       Y \\
 988921309441835904 &   110.112 &  56.0709 & 14.50 &  1.91 &  21.28 &          5.51 &       ... &             ... &  Y &         Y &         Y &         Y &        Y &       Y \\
 986459743425854848 &    110.89 &   52.684 & 12.83 &   1.2 &   13.1 &          0.96 &       ... &             ... &  Y &         Y &         Y &         N &        Y &       Y \\
 986739740933219072 &  109.5809 &  53.3539 & 11.08 &   0.8 &  11.99 &           0.4 &      2.97 &            0.04 &  Y &         Y &         Y &         N &        Y &       Y \\
 987307771832365696 &  110.1379 &   55.007 & 17.43 &  3.02 &    ... &           ... &       ... &             ... &  Y &         Y &         Y &         Y &        Y &       N \\
 989512434381005824 &   110.816 &  57.7332 & 12.84 &  1.21 &   9.24 &          1.08 &      6.66 &            0.26 &  Y &         Y &         Y &         N &        Y &       Y \\
 989412000865832448 &  111.7238 &  57.1391 &  8.75 &  0.28 &  11.88 &          1.35 &       ... &             ... &  Y &         Y &         Y &         N &        Y &       Y \\
 989512395724905600 &  110.8267 &   57.736 & 12.69 &  1.16 &   7.71 &          1.51 &      4.65 &            0.09 &  Y &         Y &         Y &         N &        Y &       Y \\
 986413254699897728 &  109.9014 &  52.3557 & 13.92 &  1.63 &   8.85 &          2.33 &      7.35 &            0.22 &  Y &         Y &         Y &         N &        Y &       Y \\
 986431224843086464 &  111.0703 &  52.2047 & 14.69 &  1.93 &   9.09 &          5.88 &     10.56 &            0.64 &  Y &         Y &         Y &         N &        Y &       Y \\
 989790812685221888 &  108.5185 &  56.9294 & 17.42 &  3.02 &    ... &           ... &       ... &             ... &  Y &         Y &         Y &         N &        Y &       Y \\
 987123156958485504 &  110.9397 &  54.8177 & 17.51 &  3.02 &    ... &           ... &       ... &             ... &  Y &         Y &         N &         Y &        Y &       Y \\
 986964002649554304 &   111.014 &  53.8388 & 18.02 &  2.92 &    ... &           ... &       ... &             ... &  Y &         Y &         Y &         Y &        Y &       N \\
 987354745890145152 &  109.8882 &  55.0877 & 11.61 &  0.98 &   7.38 &          0.86 &       ... &             ... &  Y &         Y &         Y &         Y &        Y &       N \\
 988897257625136128 &  110.0824 &  55.6896 & 16.05 &  2.75 &    ... &           ... &       ... &             ... &  Y &         Y &         Y &         Y &        Y &       N \\
 988850352285924608 &  110.8804 &  55.6272 & 18.06 &  3.05 &    ... &           ... &       ... &             ... &  Y &         Y &         Y &         Y &        Y &       N \\
 987395049863594752 &  110.0758 &  55.5564 & 10.15 &  0.63 &  10.41 &          0.25 &       ... &             ... &  Y &         Y &         N &         Y &        Y &       N \\
 987047982146021760 &  111.9727 &  54.3587 & 16.10 &  2.51 &    ... &           ... &      1.89 &            0.02 &  Y &         Y &         Y &         N &        Y &       N \\
 989713915590604800 &  109.8437 &  56.6631 & 18.09 &  3.18 &    ... &           ... &       ... &             ... &  Y &         Y &         N &         Y &        Y &       N \\
 986357626283524352 &  110.5341 &  52.2883 &  9.56 &  0.51 &  11.29 &          0.46 &       ... &             ... &  Y &         Y &         Y &         N &        N &       Y \\
 988557508531634816 &   112.372 &  54.8605 & 15.05 &  2.08 &  10.76 &          5.63 &       ... &             ... &  Y &         Y &         Y &         N &        Y &       N \\
 985556357184385664 &  113.2774 &  54.5785 &  9.85 &  0.63 &   8.24 &          2.42 &       ... &             ... &  Y &         Y &         Y &         N &        Y &       N \\
 988934572301472000 &  110.3769 &  56.1463 & 16.59 &  2.65 &    ... &           ... &      1.44 &            0.01 &  Y &         Y &         N &         Y &        Y &       N \\
 987326227307654016 &  110.7119 &  55.2842 &  6.85 & -0.05 &    6.6 &         24.05 &       ... &             ... &  Y &         Y &         N &         Y &        Y &       N \\
 989007483666095616 &  111.7559 &  56.5209 & 17.72 &   2.9 &    ... &           ... &       ... &             ... &  Y &         Y &         Y &         N &        Y &       N \\
 984018788955861248 &  112.8613 &  54.1168 & 17.75 &  2.99 &    ... &           ... &       ... &             ... &  Y &         Y &         Y &         N &        Y &       N \\
 988571041973144320 &   112.096 &  54.8537 & 18.27 &  3.32 &    ... &           ... &       ... &             ... &  Y &         Y &         N &         Y &        Y &       N \\
 986357729362738176 &  110.5321 &  52.3014 & 17.28 &  2.87 &    ... &           ... &       ... &             ... &  Y &         Y &         Y &         N &        N &       Y \\
 986768706192659712 &   109.455 &  53.5918 & 17.02 &  3.02 &    ... &           ... &       ... &             ... &  Y &         Y &         Y &         N &        Y &       N \\
 988153124475312512 &  109.3633 &  55.5093 & 18.53 &  2.96 &    ... &           ... &       ... &             ... &  Y &         Y &         N &         Y &        Y &       N \\
 989511502372001280 &  110.9872 &  57.8126 & 18.05 &   3.5 &    ... &           ... &       ... &             ... &  Y &         Y &         Y &         N &        Y &       N \\
 988952847386239744 &  109.8828 &  56.3699 & 18.41 &  3.52 &    ... &           ... &       ... &             ... &  Y &         Y &         N &         Y &        Y &       N \\
 987121851288453248 &  111.0469 &  54.7678 & 19.59 &  2.99 &    ... &           ... &       ... &             ... &  Y &         Y &         N &         Y &        Y &       N \\
 989712476778038272 &  109.8794 &  56.5813 & 14.42 &  2.15 &  13.98 &          6.78 &       ... &             ... &  Y &         Y &         Y &         N &        Y &       N \\
 989042182705921664 &  110.9134 &  56.5252 & 18.36 &  3.27 &    ... &           ... &       ... &             ... &  Y &         Y &         Y &         N &        Y &       N \\
 976696316393172736 &  110.3851 &  49.2506 & 12.53 &  1.14 &  12.57 &          0.98 &       ... &             ... &  Y &         Y &         Y &         N &        N &       N \\
 976981850114456448 &  112.3734 &  50.2457 & 10.66 &  0.71 &  12.55 &          0.63 &      2.25 &            0.02 &  Y &         Y &         Y &         N &        N &       N \\
 989322798689263616 &  113.0088 &  57.3547 & 17.37 &  2.97 &    ... &           ... &       ... &             ... &  Y &         Y &         Y &         N &        N &       N \\
 989937911021233920 &  108.7335 &  57.8749 & 15.44 &   2.2 &  51.24 &          9.73 &      2.89 &            0.03 &  Y &         Y &         Y &         N &        N &       N \\
 990128435770001664 &  108.4051 &  57.9544 &  8.72 &  0.19 &   7.45 &          0.46 &       ... &             ... &  Y &         Y &         Y &         N &        N &       N \\
 980366353064094848 &  109.4387 &  51.6512 &  8.04 &  0.02 &  13.47 &          0.18 &       ... &             ... &  Y &         Y &         Y &         N &        N &       N \\
 983783424748667264 &  112.6055 &  52.5546 & 17.88 &   3.0 &    ... &           ... &       ... &             ... &  Y &         Y &         Y &         N &        N &       N \\
 983412137711809536 &  111.9661 &  51.8702 & 17.78 &  2.95 &    ... &           ... &       ... &             ... &  Y &         Y &         Y &         N &        N &       N \\
 986880680284157056 &  111.4033 &  53.3685 & 17.94 &  2.98 &    ... &           ... &       ... &             ... &  Y &         Y &         Y &         N &        N &       N \\
 989370799243621120 &  113.1236 &  57.7923 & 17.27 &  2.91 &    ... &           ... &      1.77 &            0.02 &  Y &         Y &         Y &         N &        N &       N \\
 983609598831662720 &  113.6082 &  52.6793 & 17.93 &  3.21 &    ... &           ... &       ... &             ... &  Y &         Y &         Y &         N &        N &       N \\
 989334210417772288 &  112.9349 &  57.3629 & 15.28 &   2.2 &  19.76 &          11.9 &      2.92 &            0.03 &  Y &         Y &         Y &         N &        N &       N \\
 989314350488989056 &  113.2808 &  57.0562 & 11.29 &  0.85 &   9.48 &          0.29 &      4.75 &            0.08 &  Y &         Y &         Y &         N &        N &       N \\
 988576372027977856 &  112.7185 &  54.8564 & 16.84 &  2.93 &    ... &           ... &       ... &             ... &  Y &         Y &         Y &         N &        N &       N \\
 988554175637017088 &  112.4194 &  54.7358 & 17.89 &  3.11 &    ... &           ... &       ... &             ... &  Y &         Y &         Y &         N &        N &       N \\
 988255898748953600 &  108.0255 &  56.4503 & 16.04 &  2.52 &    ... &           ... &       ... &             ... &  Y &         Y &         Y &         N &        N &       N \\
 982984045434527744 &  112.7625 &  50.2847 & 11.71 &  0.93 &  14.33 &          1.11 &      5.37 &            0.17 &  Y &         Y &         Y &         N &        N &       N \\
 989254251010301312 &  114.0722 &  56.9368 & 17.60 &   3.0 &    ... &           ... &       ... &             ... &  Y &         Y &         Y &         N &        N &       N \\
 989327201031014784 &  113.3174 &  57.4696 & 16.51 &  2.67 &    ... &           ... &      3.31 &            0.07 &  Y &         Y &         Y &         N &        N &       N \\
 986143320298473344 &  114.0672 &   56.381 &  8.23 &  0.12 &   7.71 &          0.39 &       ... &             ... &  Y &         Y &         Y &         N &        N &       N \\
 980282137345565184 &  110.5529 &  51.1837 & 15.42 &  2.13 &    ... &           ... &       ... &             ... &  Y &         Y &         Y &         N &        N &       N \\
 977211785483558656 &  110.9881 &  50.2171 & 17.19 &  2.94 &    ... &           ... &       ... &             ... &  Y &         Y &         Y &         N &        N &       N \\
 988619901520446336 &  111.9144 &  54.9836 & 19.78 &  2.89 &    ... &           ... &       ... &             ... &  Y &         Y &         N &         N &        Y &       N \\
 989766455926914432 &  108.3799 &  56.8396 & 16.61 &  3.03 &    ... &           ... &       ... &             ... &  Y &         Y &         Y &         N &        N &       N \\
 989541468360142208 &  112.5391 &   57.885 & 18.01 &  3.22 &    ... &           ... &       ... &             ... &  Y &         Y &         Y &         N &        N &       N \\
 988781052989337472 &  111.9967 &  56.0498 & 18.76 &  3.05 &    ... &           ... &       ... &             ... &  Y &         Y &         N &         N &        Y &       N \\
 983821396555375488 &  112.7095 &  52.9595 & 16.03 &  2.62 &    ... &           ... &       ... &             ... &  Y &         Y &         Y &         N &        N &       N \\
 989418696718762368 &  111.4899 &  57.2837 & 18.62 &  3.46 &    ... &           ... &       ... &             ... &  Y &         Y &         N &         N &        Y &       N \\
 989426530738246016 &  112.0985 &  57.2512 & 13.40 &  1.46 &   8.82 &          2.86 &       ... &             ... &  Y &         Y &         Y &         N &        N &       N \\
 982983431255441024 &  112.7668 &   50.246 & 12.42 &  1.17 &   8.89 &          7.62 &       ... &             ... &  Y &         Y &         Y &         N &        N &       N \\
 989052112670442496 &  110.3316 &   56.583 & 19.64 &  3.21 &    ... &           ... &       ... &             ... &  Y &         Y &         N &         N &        Y &       N \\
 980635355455537152 &  108.6972 &  52.4191 & 16.34 &  2.66 &    ... &           ... &       7.3 &            0.31 &  Y &         Y &         Y &         N &        N &       N \\
 988682131302720896 &   112.911 &  55.2023 & 17.98 &  3.18 &    ... &           ... &       ... &             ... &  Y &         Y &         Y &         N &        N &       N \\
 989386437219054208 &  111.9204 &  56.7997 & 19.71 &  2.88 &    ... &           ... &       ... &             ... &  Y &         Y &         N &         N &        Y &       N \\
 989439316857826304 &  111.7266 &  57.3529 & 18.03 &  3.12 &    ... &           ... &       ... &             ... &  Y &         Y &         Y &         N &        N &       N \\
 990028032321723136 &  107.1289 &  57.2901 & 16.57 &  2.56 &    ... &           ... &       ... &             ... &  Y &         Y &         Y &         N &        N &       N \\
 987116907781125248 &  111.1088 &  54.6653 & 17.21 &  2.98 &    ... &           ... &       ... &             ... &  Y &         N &         Y &         N &        N &       N \\
 975786711039098752 &  115.1815 &  48.0467 & 18.22 &  3.21 &    ... &           ... &       ... &             ... &  Y &         Y &         N &         N &        N &       N \\
 976398554900260992 &  110.9972 &  48.3394 & 11.93 &  0.97 &  14.05 &          0.94 &      5.35 &             0.1 &  Y &         Y &         N &         N &        N &       N \\
 934445662246052992 &  117.3715 &   50.388 & 19.32 &  2.99 &    ... &           ... &       ... &             ... &  Y &         Y &         N &         N &        N &       N \\
 976427971132363264 &  111.4665 &  48.5476 &  6.75 & -0.06 &   1.81 &          6.29 &       ... &             ... &  Y &         Y &         N &         N &        N &       N \\
 981409510425138944 &  107.9282 &  52.7908 & 18.73 &  3.21 &    ... &           ... &       ... &             ... &  Y &         Y &         N &         N &        N &       N \\
 990715128302384896 &  109.2736 &  59.8146 & 11.88 &  0.95 &   7.99 &           0.6 &       ... &             ... &  Y &         Y &         N &         N &        N &       N \\
 987406869612182016 &  109.6723 &  55.5808 & 19.63 &  3.15 &    ... &           ... &       ... &             ... &  Y &         Y &         N &         N &        N &       N \\
 977250371469717504 &  111.0723 &  50.5975 & 14.88 &  2.05 &  15.95 &          5.74 &       ... &             ... &  Y &         N &         Y &         N &        N &       N \\
1085819516851255680 &  113.2548 &  59.4704 & 16.47 &  2.53 &    ... &           ... &       ... &             ... &  Y &         Y &         N &         N &        N &       N \\
 993222805088975104 &  104.2593 &  52.6283 & 18.01 &  3.28 &    ... &           ... &       ... &             ... &  Y &         Y &         N &         N &        N &       N \\
 986815641595140992 &  108.5942 &   53.898 & 14.08 &  1.62 &  -9.99 &          2.14 &       ... &             ... &  Y &         N &         Y &         N &        N &       N \\
 983396778909086976 &  112.5652 &   52.173 & 16.71 &  2.69 &    ... &           ... &       ... &             ... &  Y &         Y &         N &         N &        N &       N \\
1081800591396594432 &  119.0079 &   57.476 & 17.29 &  2.79 &    ... &           ... &       ... &             ... &  Y &         Y &         N &         N &        N &       N \\
 983196594776807808 &  111.8892 &  50.6122 & 18.63 &  3.42 &    ... &           ... &       ... &             ... &  Y &         Y &         N &         N &        N &       N \\
 987952566682056320 &  106.2823 &  55.4264 & 18.72 &  3.48 &    ... &           ... &       ... &             ... &  Y &         Y &         N &         N &        N &       N \\
 976063994127719296 &   114.221 &  48.8949 & 17.68 &  3.24 &    ... &           ... &       ... &             ... &  Y &         Y &         N &         N &        N &       N \\
 989661723148236160 &  111.4738 &  58.6683 & 19.96 &  2.42 &    ... &           ... &       ... &             ... &  Y &         Y &         N &         N &        N &       N \\
 989436563782825984 &  112.0646 &  57.4916 & 16.20 &  2.45 &    ... &           ... &       ... &             ... &  Y &         N &         Y &         N &        N &       N \\
1082193945977352448 &  118.1075 &  57.5417 & 16.01 &  2.32 &    ... &           ... &       ... &             ... &  Y &         Y &         N &         N &        N &       N \\
 986586427781001472 &  109.2919 &  52.4983 & 17.28 &  2.84 &    ... &           ... &       ... &             ... &  Y &         N &         Y &         N &        N &       N \\
 980142911685392384 &  110.3125 &  50.4801 &   15.73 &   2.43 &    ... &           ... &       ... &             ... &  N &         Y &         Y &         N &        N &       N \\
 990364860833413888 &  109.5524 &  58.7623 &   17.13 &   2.57 &    ... &           ... &       ... &             ... &  N &         Y &         Y &         N &        N &       N \\
1087349929661806080 &  110.8443 &   61.535 &   17.89 &   3.00 &    ... &           ... &       ... &             ... &  N &         Y &         Y &         N &        N &       N \\
    \enddata
    \tablecomments{
    \tablenotetext{a}{From \gaia\ DR3 \citep{GaiaCollaboration2023}}
    \tablenotetext{b}{From \cite{Boyle2026_tars}}
    \tablenotetext{c}{Identified using \texttt{FriendFinder}.}
    \tablenotetext{d}{Identified in Alessi~84 \citep{Hunt2024}.}
    \tablenotetext{e}{Identified in Theia~214 \citep{Kounkel2019}.}
    \tablenotetext{f}{Identified in CWNU~1128 \citep{He2022}.}
    \tablenotetext{g}{Identified in OCSN~304 \citep{Qin2023}.}
    \tablenotetext{h}{Identified in UPK~343 \citep{Sim2019}.}
    }
    \end{deluxetable*}
\end{longrotatetable}

\bibliographystyle{aasjournal.bst}
\bibliography{Mannbib, planetSearch}

\end{document}